\documentclass{article}
\usepackage{arxiv}
\usepackage[utf8]{inputenc} 
\usepackage[T1]{fontenc}    
\usepackage{url}            
\usepackage{booktabs}       
\usepackage{amsfonts}       
\usepackage{nicefrac}       
\usepackage{microtype}      
\usepackage{times}
\usepackage{float}
\usepackage{booktabs}
\usepackage{textcomp}
\usepackage{amsmath}
\usepackage[flushleft]{threeparttable}
\usepackage{fancyhdr,graphicx,amssymb}
\usepackage[ruled,vlined]{algorithm2e}
\usepackage{fullpage}
\usepackage{times}
\usepackage{float}
\usepackage[font=small,skip=5pt]{caption}
\usepackage[justification=centering]{caption}
\usepackage{booktabs}
\usepackage{textcomp}
\usepackage[flushleft]{threeparttable}
\usepackage[ruled,vlined]{algorithm2e}
\usepackage{multirow}
\usepackage{subcaption}
\usepackage{caption}
\usepackage{float}
\makeatletter
\newcommand{\setlabel}[1]{\edef\@currentlabel{#1}\phantomsection\label}
\makeatother
\usepackage[colorlinks=true]{hyperref}
\usepackage{nameref}

\usepackage{expl3}

\ExplSyntaxOn
\newcommand\latinabbrev[1]{
  \peek_meaning:NTF . {
    #1\@}%
  { \peek_catcode:NTF a {
      #1.\@ }%
    {#1.\@}}}
    
\providecommand{\keywords}[1]{\textbf{\textit{Keywords---}} #1}

\newcommand{\ignore}[1]{}
\ExplSyntaxOff

\def\eg{\latinabbrev{e.g}}
\include{pythonlisting}
\graphicspath{ {./figures/} }
\title{Mining Temporal Evolution of Knowledge Graphs and Genealogical Features for Literature-based Discovery Prediction}

\author{
   Nazim Choudhury \\
   Department of Computer Science and Engineering\\
   University of South Florida\\
   Tampa, FL 33620 \\
  \texttt{nachoudhury@usf.edu} \\
   \And
    Fahim Faisal\thanks{Corresponding Author.} \\
    Dept. of Computer Science and Engineering\\
    Islamic University of Technology\\
    Dhaka, Bangladesh\\
    \texttt{fahimfaisal@iut-dhaka.edu} \\
    \And
    Matloob Khushi\\
    School of Computer Science\\
    The University of Sydney\\
    Sydney, NSW 2006, Australia\\
    \texttt{matloob.khushi@sydney.edu.au} \\
}

\begin{document}
\maketitle

\begin{abstract}
Literature-based discovery process identifies the important but implicit relations among information embedded in published literature. Existing techniques from Information Retrieval (IR) and Natural Language Processing (NLP) attempt to identify the hidden or unpublished connections between information concepts within published literature, however, these techniques overlooked the concept of predicting the future and emerging relations among scientific knowledge components encapsulated within the literature. Keyword Co-occurrence Network (KCN), built upon author selected keywords (i.e., knowledge entities), is considered as a knowledge graph that focuses both on these knowledge components and knowledge structure of a scientific domain by examining the relationships between knowledge entities. Using data from two multidisciplinary research domains other than the medical domain, capitalizing on bibliometrics, the dynamicity of temporal KCNs, and a recurrent neural network, this study develops some novel features to successfully predict the future literature-based discoveries - the emerging connections among knowledge units. 
Temporal importance extracted from both bipartite and unipartite networks, communities defined by genealogical relations, and the relative importance of temporal citation counts were used in the feature construction process. 
Both node and edge-level features were input into a recurrent neural network to forecast the feature values and predict the future relations between different scientific concepts represented by the author selected keywords. 
High performance rates suggest that these features are supportive both in predicting the future literature-based discoveries and emerging trend analysis.
\end{abstract}


\section {Introduction}
Statistical bibliography or bibliometrics ~\cite{pritchard1969statistical} has been supporting researchers to address the challenges related to the rapid growth of scholarly publications and scientific knowledge.
Employing two network-based methods (i.e., co-citation and keyword co-occurrences network), bibliographic coupling ~\cite{kessler1963bibliographic} method in bibliometrics enabled us to explore the structure of scientific and technical knowledge. 
While the co-citation focuses on the structure of scientific communication by analyzing citation links, keyword co-occurrence network (KCN) or co-word network, focuses on knowledge components and knowledge structure by examining co-appearances of keywords found in the literature.
Therefore, it is also known as knowledge graph ~\cite{popping2003knowledge}.
Associated as metadata information within the scholarly publications, the author selected keywords are considered as the carriers of knowledge units, or knowledge entities ~\cite{su2010mapping}. 
These hand-picked signal words/terms/themes/conceptual keywords denote authors’ understandings of their work, the thematic context of their research, and facilitate knowledge discovery ~\cite{song2013discovering}.
Further, these topical concepts or themes are also used for indexing purpose in digital libraries.
The co-appearance of two author selected keywords in an article defines a certain intrinsic relationship between them whereas multiple such instances denote the strength of their relationships ~\cite{yang2011integration}.
Representing such co-occurring relationships between the knowledge entities, KCN is bears both theoretical and practical implications including literature-based discovery ~\cite{choudhury2016time}.

Literature-based discovery ~\cite{kostoff2005method}, widely spelled as Literature Based Discovery (LBD), is the process that seeks to discover new knowledge (innovation) from the information embedded in published literature.
In both automated and semi-automated ways, it identifies the hidden (implicit) and important connections among knowledge components. 
Rapid growth of scholarly publications and associated scientific knowledge prompted scholars to restrict themselves within their narrow specialties (i.e., fragments within a broad domain) and cite only from the related articles ~\cite{ganiz2005recent}. 
Consequently, useful connections between fragmented information remain unnoticed due to the lack of awareness of work from mutually exclusive fragments. 
LBD addresses this problem of knowledge overspecialization and strives to identify the implicit and novel connections from the concepts explicitly published in the scientific literature. 
Preiss et al. ~\cite{preiss2015exploring} reported different applications of this approach including identification of treatments for diseases, drug re-purposing, disease candidate gene discovery, and drug side effect prediction. 
Similarly, Henry et al. ~\cite{henry2017literature} reported application areas outside the biomedical domain such as efficient water purification systems, development acceleration of developing countries, potential bio-warfare agents categorization, climate change studies, and identification of promising research collaborations.

Most methodologies addressing LBD are based on  Swanson's `ABC' model ~\cite{swanson1986fish} and predominantly rely on text analysis ~\cite{swanson1997interactive}. 
Further, most studies in LBD used MEDLINE as the literature database.
Sebastian et al. ~\cite{sebastian2017emerging} divided existing techniques of literature-based knowledge discovery from scientific literature into two categories: (i) traditional, and (ii) emerging approaches. 
The traditional approach dominates the current research landscape.
These are mainly knowledge-based and comprised of lexical statistics or graph-theoretic methods that require domain knowledge. 
On the other hand, emerging approaches prompt new trends and unprecedented paradigm shifts in the knowledge discovery process.
These methods consider the occurrence and co-occurrence frequencies of keywords, probability distributions, association rule mining, graph data structure, temporal features, relational attributes, and supervised classification approach.
A literature review on these methods is presented in the ~\ref{sec:Related works} section.
Being a knowledge network representation technique, KCN inherently fits best in the LBD process.
However, despite their success in mapping scientific knowledge structure, the evolutionary dynamics of KCN, and metadata information (\eg authors, citations) associated with scholarly articles  ~\cite{ding2013entitymetrics} are underutilized in the existing LBD techniques.
Temporal changes of metadata information, and communities of keywords extracted from dynamic KCN can also be considered as non-trivial factors than mere frequencies or connectivity information in predicting the underlying complex and implicit relations between keywords.
Most importantly, existing methods in LBD concentrates on identifying the implicit or undisclosed relations between keywords in the published literature instead of predicting their future relations those are yet to emerge.
This can facilitate emerging trend analysis which can be supportive for both researchers and science policymakers.
To this end, by formulating the LBD prediction problem as a dynamic supervised link prediction problem, we propose a recurrent neural network (RNN)-based method to predict the emerging associations (i.e., co-occurrences) between author selected keywords found in the scientific literature.
This prediction process incorporated features constructed by leveraging the temporal dynamics of KCNs, evolutionary metadata information (authors, citation) and genealogical communities of keywords
Due to the temporal nature of the constructed features and binary supervised classification task, we considered the Long Short Term Memory (LSTM) - an artificial RNN architecture, suitable for sequence classification.
The research objective of this study is to \textit{\textbf{ predict the co-evolution of knowledge entities (author selected keywords) found in the scientific literature}} to facilitate the prediction of future LBD. 
Since the prediction problem was framed as a supervised learning process, the scientific contribution of this study is the development of representative features to describe and successfully classify instances (non-connected keyword pairs). 
By considering dynamic KCNs, this study \textit{\textbf{ develops some novel features by leveraging the temporal dynamics of KCNs, metadata information (i.e., authors, articles, citations), and finally, communities defined via deriving ancestral lineages of keywords extracted from the dynamic KCNs }}.
The prediction performances of the proposed features, applied over scholarly datasets on two topics \footnote{datasets and codes are available in https://github.com/faisal-iut} ( i.e., \textbf{\textit{Obesity, Sleep Apnea}}),  are also reported. 
These generic features are domain independent and can be applied to any multidisciplinary scientific disciplines to predict the emerging trends.
\section*{Related Works}\setlabel{Related Works}{sec:Related works}

Literature-based discovery (LBD) identifies the implicit relations from the explicit information.
It is principally used in biomedical research where running experiments is expensive. 
This knowledge discovery process broadly encompasses lexical and semantic text analyses of articles found in the MEDLINE database. 
Most of the LBD techniques employed frequency-based approaches.
The underlying assumption is that discoveries are likely to arise from logical connections among source, intermediate, and target concepts (keywords) based on either their frequent/infrequent  (co)occurrences in the literature, or common/rare connections to a knowledge base ~\cite{cameron2015context}.
Gordon and Dumais ~\cite{gordon1998using} took the advantages of both co-occurrence frequencies and Latent Semantic Indexing (LSI) used in analyzing relationships between a set of documents.
Hristovski et al ~\cite{hristovski2001supporting} used association rule mining and Unified Medical Language Systems (UMLS) to discover the relationships between medical concepts.
Yetisgen-Yildiz and Pratt developed ~\cite{yetisgen2006using} `LitLinker' that incorporated knowledge-based methodologies with statistical approach considering the background distribution of keyword probabilities.
With the help of Fuzzy Set Theory and probabilistic model of relationships, Wren et al. ~\cite{wren2004knowledge}  developed a computational method to identify large sets of relationships between unrelated items within scientific reports.

Ensemble-based approaches combined both statistical and temporal features to find the intermediate keywords connecting both source and target keywords.
These approaches were explored to find meaningful links between two disparate sets of articles in MEDLINE  ~\cite{smalheiser2009arrowsmith}.
Relational techniques ~\cite{ahlers2007using, hu2005semantic} used the explicit semantic relationships (i.e., predicates) between concepts where such predicates were typically obtained from structured background knowledge or known \textit{a priori} via domain experts.
Few approaches focused on graph data structures to discover knowledge from the literature by creating subgraphs based on the binary relationships between literary concepts. 
These relationships were also drawn from semantic predications which were extracted directly from assertions in MEDLINE literature. 
By weighting links using degree centrality, Wilkowski et al. ~\cite{wilkowski2011graph} developed such a graph-theoretic approach.
Their approach used an iterative greedy strategy to create the `best’ subgraph  with the help of semantic predications. 
Cameron et al. ~\cite{cameron2015context} also used semantic predication graph and introduced a method to automatically find clusters of contextually similar paths in the graph. 
These clusters were then used to identify the latent associations between disjoint concepts in the literature to reconstruct eight scientific discoveries.
Like most LBD methods, these graph-based data structures were also primarily constructed in the  biomedical domain using predicative relations extracted from MEDLINE literature.

Contemporary approaches developed heterogeneous networks capable of encoding richer information and better semantics between various real world objects ~\cite{sun2012mining}. 
Known as Heterogeneous Bibliographic Information Network (HBIN), these networks represent a collection of scientific publications as a network of heterogeneous bibliographic objects (e.g., keywords, authors etc.).
HBIN allowed various information to flow across different types of objects and links to capture the previously unknown associations between research articles.
It also harnessed the meta-path features found in HBIN networks to discover the latent associations. 
Sebastian et al. ~\cite{sebastian2017learning} used lexico-citation features of HBIN networks to predict the co-citation links between articles from previously disconnected research areas.
Similarly, Ding et al. ~\cite{ding2013entitymetrics} used information from the literature in the form of an `entitymetrics' graph.
The objective was to model the latent relationships among biomedical entities (e.g. diseases, drugs) based on the existing citation relationships among the corresponding articles. 
Apart from predicting the implicit and/or hidden relationships between disjoint sets of articles, researchers  ~\cite{ren2014cluscite, liu2014full} also used HBIN to predict the citation count.
HBIN-based LBD used simple statistics and does not require sophisticated and domain-specific NLP tools and ontologies. 
It also facilitated the utilization of metadata information in constructing features for prediction task. 
However, to predict future links between author selected keywords instead of links between articles, processing HBIN and calculating meta-path features will be computationally intensive.  

One of the emerging approaches ~\cite{kastrin2016link} conjectured the task of predicting implicit relationships as a classification task by leveraging the link prediction methodology of network science.
It describe the associations between different concepts/keywords/topics using networks where the links represent their semantic or co-occurrence relationships.
This approach was primarily used in predicting the implicit links within a co-occurrence network of Medical Subject Headings (MeSH) terms.
Crichton et al. ~\cite{crichton2018neural} recently investigated how inputs from four node representation algorithms affect the performance of a neural link predictor on random- and time-sliced biomedical graphs (i.e., Drug-Target Interactions, Protein-Protein Interaction and LBD)  containing information relevant to drugs, protein and literature.
Katukuri et al. ~\cite{katukuri2012hypotheses} used manually-created features in a supervised link prediction task on a large-scale biomedical network of concepts co-occurrences.
The objective was to predict links those represented scientific hypotheses in a time-sliced corpus. 
The authors extracted relevant information from the biomedical corpus to generate a concept network and concept-author map.
They also developed a set of heterogeneous features by exploiting random walk, network neighborhood and common authorship.
Wang and Zeng ~\cite{wang2013predicting} used two-layer graphical model, called restricted Boltzmann machine, to perform similar link prediction task on Drug-Target Interactions (DTI) network.
Their objective was to predict multiple types of DTIs,  unknown drug-target relationships or drug's modes of action.
However, Lu et al. ~\cite{lu2017link} pointed out the limitation of such machine learning models in DTI predictions despite their high performance.
This included the absence of additional information about the characteristics of drugs, targets and DTIs, (e.g.,chemical structure, genome sequence, binding types, causes of interactions, etc.).
Therefore, the authors used topological similarity indices, such as Common Neighbours and Katz ~\cite{katz1953new} index from complex network theory, to predict links in a DTI network.
Our study is probably closer to this set of machine-learning based studies however it differs in different ways which is described later in this section.

In summary, existing research used combinations of lexical distributional statistics, graph-theoretic measures,  heterogeneous network-based methods, and machine learning models.
Predominantly capitulated on MEDLINE database, most existing methods require domain knowledge to interpret relations between knowledge entities (drug, target, protein, gene etc.).
Graph-theoretic methods considered static networks which is unable to capture the temporal aspects of network evolution such as `recency', genealogical traits (origins of relationships), and time-variant frequencies crucial for predicting the emerging associations in evolving networks.
Despite the merits of existing approaches, there exists lack of generalized predictive features applicable to any research domain.
These features should be free from any domain-specific knowledge-base for predicative relationships or relax the requirements of domain experts for interpretation.
It should accommodate metadata information which is commonly associated with any scientific literature like authors, keywords, affiliation etc, and integrate the temporal information since scientific knowledge structure is inherently dynamic.
These facts are the motivation behind this study.
Further, in most cases, these methods focused only on identifying implicit(hidden) relationships between scientific concepts which are yet to be recognized from their explicit relations instead of predicting the future relations those are yet to emerge.
To this end, this study develops a recurrent neural network-based method to predict emerging LBD instead of identifying the hidden relations between keywords/concepts.
It manifests the knowledge evolution as a dynamic process and leverages the evolutionary aspects of knowledge graphs, temporal recency and citation information to develop hand-crafted features suitable for the prediction task.
\section {Scholarly Datasets}
\setlabel{Scholarly Datasets}{sec:data_acquisition}
We extracted our scholarly datasets on two topics and the source of these datasets is  `Scopus' -the largest abstract and citation database of peer-reviewed literature. 
The first search keyword was \textbf{`sleep apnea'}, also spelled as sleep apnoea, and the second keyword was \textbf{`obesity'}. 
The first topic is related to the serious sleep disorder that occurs when a person's breathing is interrupted or a person experiences periods of shallow breathing during sleep. 
The second topic is related to overweight and represents a complex disorder involving the accumulation of excessive body fat to an extent such that it may have negative impact on health of individuals.
Article metadata information including publication year, article title, authors, affiliations, author selected keywords and citation count were extracted from the Scopus digital library by considering the following constraints: (i) article published in English journals, (ii) the search keywords are present in the article's titles and abstracts, and (iii) articles published within the duration 2007-2016.
For the sake of brevity, $G_s$ and $G_o$ will be used to denote the dataset related to sleep apnea and obesity respectively for the rest of the article. 

\begin{table}[!h]
\centering
\caption{Statistics of two datasets}
\label{table:data}
\begin{tabular*}{\textwidth}{l@{\extracolsep{\fill}}rrr}
\toprule
Dataset	&  \# Articles 	&  \# Keywords	& Duration \\
\midrule
$G_{s}$    	&   29203 			&    12721 	& 2007-2016	\\
$G_{o}$     	&   107745 			&    11643 	& 2007-2016	\\
\bottomrule
\end{tabular*}
\end{table}
The author selected keywords are crucial to know about the thematic context, topics and related concepts of the corresponding scholarly articles.
Recent advances in network science  ~\cite{borner2007network} have prompted researchers to address the mapping and understanding of scientific knowledge evolution via different types of bibliometric networks.
These networks consist of nodes representing different scholarly items like publications, journals, researchers, or keywords and edges indicating the relations between pairs of nodes. 
According to Van Eck and Waltman ~\cite{van2014visualizing}, the most commonly studied types of relations are: (i) citation relations, (ii) keyword co-occurrence relations, and (iii) co-authorship relations. 
To predict the emerging relations between different scientific concepts/keywords, we took the advantage of temporal keyword co-occurrence relations and constructed dynamic KCNs in both $G_s$ and $G_o$ for each year. 
Therefore, as an integral part of dynamic KCN construction. we extracted author selected keywords appeared in more than one articles. 
Keywords failed to gain such minimum attention from the research community were considered to be irrelevant to the corresponding research and thus discarded.
Keyword extraction phase was followed by the text processing phase that includes text cleaning and transformation. 
Authors used different spellings and acronyms to represent their keywords those represent same semantic meaning. 
Firstly, any unwanted space and other discriminators were eliminated from the keyword list with the help of NLTK text pre-processing tools ~\cite{Bird:2009:NLP:1717171}. 
This step also included lower-casing all keywords, removing singular-plural differences (\eg, epidemiological studies $\leftrightarrow$ epidemiological study, dilator muscles $\leftrightarrow$ dilator muscle) and lemmatization of some commonly used keywords (\eg, dreaming $\leftrightarrow$ dream). 
Secondly, semantically related common keywords were uniformly presented by using one representative word for all (\eg, aspect $\leftrightarrow$ feature). 
Most widely used abbreviated keywords were kept unchanged  however others were changed to it's full forms (\eg,  bmi $\leftrightarrow$ body mass index).
There were some abbreviations with different full forms (\eg, egfr $\leftrightarrow$ (estimated glomerular filtration rate, epidermal growth factor receptor)). 
In these cases, the content of the corresponding article was verified to identify the right full form.
Besides, all numbers and their corresponding roman forms were used in number format. 
Basic dataset statistics including the number of these cleaned and transformed keywords are presented in Table \ref{table:data}.
It is noteworthy that although the data collection period includes 2007, however, in the experiment, we considered 2008-2014 duration as our actual training period.
This fact will be explained in the later section.

\begin{table}[!h]
\centering
\captionsetup{justification=centering}
\caption{Yearly statistics of nodes and edges in keyword co-occurrence networks (KCNs) in\\ $G_s$=Sleep Apnea and $G_o$=Obesity}
\label{table:node-edge state}

\begin{tabular*}{\textwidth}{c@{\extracolsep{\fill}}rrrrrrrrr}
\hline
 &  & 2007 & 2008 & 2009 & 2010 & 2011 & 2012 & 2013 & 2014  \\ \hline
\multirow{2}{*}{$G_s$} & Nodes & 388 & 459 & 470 & 496 & 536 & 584 & 634 & 583 \\ \cline{2-10} 
 & Edges & 689 & 773 & 688 & 786  & 977 & 1054 & 1280 & 1144 \\ \hline
\multirow{2}{*}{$G_o$} & Nodes & 2109 & 2356 & 2092 & 2298 & 2408 & 2579 & 2582 & 3217 \\ \cline{2-10} 
 & Edges & 9569 & 10282 & 10477 & 12001 & 14271 & 16158 & 15851 & 18238 \\ \hline
\end{tabular*}
\end{table}
\section{Background}
\setlabel{Background}{subsec:bckgnd}
\subsection {Knowledge Evolution and Dynamic KCN}
\setlabel{Keyword co-occurrence network}{subsec:kcn}
Currently, scientific knowledge creation is  dynamic and interdisciplinary in nature where different avenues of research converge and new connections emerge among disjoint and existing areas of science ~\cite{pan2012evolution}.
This knowledge is generally  incremental except few revolutionary and fundamental changes.
The development and evolution of science and technology create new knowledge from the previously accumulated and ubiquitous information ~\cite{lee2009approach}.
New hypotheses are being postulated by encompassing existing scientific concepts from multiple domains.
Agusti Canals ~\cite{canals2005knowledge} pointed out that the diffusion of scientific knowledge can be mapped into a network structure where knowledge propagates via interactions among networked agents.  
A knowledge graph (network) is a specific kind of knowledge representation technique that uses a semantic network structure where nodes are keywords and links represent their causal relationships ~\cite{popping2003knowledge}
Statistically significant and non-trivial co-occurrence patterns of keywords not only represent their semantic affinity ~\cite{montemurro2013keywords} and relatedness ~\cite{schulz2014semantic} but also their causal relationships ~\cite{mcnorgan2007feature}.
Therefore, using the extracted, cleaned, and processed keywords, as described above, dynamic Keyword Co-occurrence Networks (KCNs) were constructed for each year in 2008-2015.
KCN is defined as an undirected network $G(V, E)$ where nodes $V$ is the set of author selected keywords and $E$ is the set of edges where each edge represents the co-appearance of two keywords in the same article. 
Multiple such co-appearances define the edge weights which is ignored in this study.
Dynamic KCNs are the temporal networks $G_t = (v_t, e_t)$ for time period $t = 1,2,...,T$ where 
$v_t$ is the set of nodes or keywords and 
$e_t$ is the set of edges connecting the set of keywords $v_t \in V$ at time $t$. 
These edges at time $t$ can be new or recurring. 
Table \ref{table:node-edge state} provides statistics of the number of nodes (keywords) and edges (co-occurrences) per year in both datasets.
\ignore{
\begin{table}[!h]
\centering
\begin{threeparttable}
\caption{Node and edge statistics of co-word networks for the two datasets}
\label{table:node-edge state}
\begin{tabular*}{\textwidth}{l @{\extracolsep{\fill}}lrrrrrr}
\toprule
Dataset name & Year	& $E_T$ &  $E_{T+1}$ & $V_T$	& Positively-labeled edges & Negative-labeled Edges 		\\
\midrule
\multirow{8}{*}{$G_{s}$} & 2008 &  773 & & 459 & \multirow{7}{*}{1441}&\multirow{7}{*}{590195}		\\
		    & 2009	&  620		&  		  & 191		&				    & 			\\
		    & 2010	&  674		&  		  & 133		&				    & 			\\
		    & 2011	&  799		&  		  & 95		&				    & 			\\
		    & 2012	&  808		&  		  & 79		&				    & 			\\
		    & 2013	&  977		&  		  & 90		&				    & 			\\
		    & 2014	&  826		&  		  & 46		&				    & 			\\\cmidrule{2-7}
		    & 2015	&  		 	&  5142	  & 1093		&				    & 			\\
\midrule

\multirow{8}{*}{$G_{o}$}& 2008	&  10285 &  & 2357& \multirow{7}{*}{14203}&\multirow{7}{*}{8033692}	\\
			& 2009	&  8894 	&  		  & 531		& 					&				\\
			& 2010	&  9210 	&  	      & 276		&				    & 			\\
			& 2011	&  9970 	& 	      & 120		&				    & 			\\
			& 2012	&  556	 	&  		  & 61		&				    & 			\\
			& 2013	&  530	 	&  		  & 28		&				    & 			\\
			& 2014	&  13892 	&  		  & 652		&				    & 			\\\cmidrule{2-7}
			& 2015	&  		 	&  21610  & 4025	&				    & 			\\
\bottomrule
\end{tabular*}
\begin{tablenotes}
\item $E_T$ represents the set of edges in the training phase (2008-2014) and $E_{T+1}$ represents edges in the test phase (2015). $E_T$ is the set of all nodes (Keywords) common between training and test years. Positively labeled edges are those non-connected keyword-pairs from training phase appeared in the test phase whereas negatively labeled edges are all other keyword-pairs non-existent in both training and test phase.
\end{tablenotes}
\end{threeparttable}
\end{table}
}
In Table \ref{table:gr_ke}, we present the basic statistics of the evolutionary patterns observed in dynamic KCN for both $G_o$ and $G_s$. 
The dynamicity of keyword co-occurrences denotes that new research topics, hypotheses, or directions emerge over time through co-appearances of existing keywords.

Three different scenarios can be observed in dynamic KCNs.
\begin{table}[!h]
\centering
\begin{threeparttable}
\caption{Evolutionary statistics of nodes (author selected keywords) and edges (co-occurrences in the same article) for two datasets in this study. $V_t$ = keywords in year $t$, $V_{n}$ = new keywords each year, $V_{o}$ = old keywords from the previous year(s), $E$ = edges, $E_{o \leftrightarrow o}$ = recurring edges between $v\in V_{o}$, $E{}'_{o\leftrightarrow o}$ = new edges between  $v\in V_{o}$, $E_{n \leftrightarrow o}$ = edges between $v\in V_{o}$ and $v\in V_{n}$, $E_{n \leftrightarrow n}$ = edges between $v\in V_{n}$. The term `old' in a particular year denotes the set of keywords appeared in previous year(s)}
\label{table:gr_ke}
\begin{tabular*}{\textwidth}{l@{\extracolsep{\fill}}rrrrrrrrr}
\toprule
{} &  year &    $V_t$ &   $V_{n}$ &   $V_{o}$ &     $E$ &  $E_{o \leftrightarrow o}$ &   $E{}'_{o\leftrightarrow o}$ & $E_{n \leftrightarrow o}$ & $E_{n \leftrightarrow n}$ \\
\midrule
$G_{s}$   &  2007 &   857 &  857 &     0 &  1486 &      0 &      0 &     0 &  1486 \\
     &  2008 &   968 &  572 &   396 &  1690 &    144 &    530 &   777 &   239 \\
     &  2009 &   978 &  438 &   540 &  1587 &    202 &    609 &   636 &   140 \\
     &  2010 &  1000 &  332 &   668 &  1648 &    264 &    800 &   483 &   101 \\
     &  2011 &  1172 &  345 &   827 &  2264 &    387 &   1166 &   599 &   112 \\
     &  2012 &  1240 &  279 &   961 &  2399 &    271 &   1524 &   531 &    73 \\
     &  2013 &  1328 &  272 &  1056 &  2715 &    424 &   1692 &   539 &    60 \\
     &  2014 &  1367 &  291 &  1076 &  2728 &    500 &   1605 &   502 &   121 \\
     &  2015 &  1262 &  157 &  1105 &  2347 &    512 &   1530 &   235 &    70 \\
\midrule
$G_{o}$  &  2007 &  3207 &  3207 &     0 &  15725 &      0 &      0 &     0 &  15725 \\
    &  2008 &  3515 &  1217 &  2298 &  16826 &   3167 &   8995 &  4336 &    328 \\
    &  2009 &  4206 &  1709 &  2497 &  21122 &   4515 &   9091 &  6695 &    821 \\
    &  2010 &  4696 &  1152 &  3544 &  24093 &   6578 &  13154 &  4094 &    267 \\
    &  2011 &  5059 &   658 &  4401 &  29261 &   6514 &  19962 &  2698 &     87 \\
    &  2012 &  5426 &   364 &  5062 &  32663 &   9944 &  20900 &  1744 &     75 \\
    &  2013 &  5418 &    77 &  5341 &  32076 &  11702 &  19987 &   382 &      5 \\
    &  2014 &  4283 &   738 &  3545 &  22410 &   6522 &  12386 &  3251 &    251 \\
    &  2015 &  4466 &   367 &  4099 &  23606 &   8043 &  13860 &  1548 &    155 \\
\bottomrule
\end{tabular*}
\begin{tablenotes}
\item 
\end{tablenotes}
\end{threeparttable}
\end{table}
As observable in Table  \ref{table:gr_ke}, firstly, new edges emerge each year when new keywords co-appear in articles.
Secondly, new keywords form edge with old (existing) keywords (appeared in previous years).
Finally, edges are formed in a year between two old keywords from the previous year(s), where these old keywords appeared in different articles but not co-appeared in the same article.
The term `old' in any particular year $t$ denotes the set of keywords appeared in year(s) prior to $t$. \eg, in $G_{s}$ for the year 2010, the number of $V_{o}$ is $668$ which denotes that out of $V_t=1000$ keywords in 2010, $668$ keywords appeared within 2007-2009. 
It is observable from the table that the number of new keywords normally decreases.
When a new hypothesis gains considerable attentions in the subsequent years, the related keywords become significant.
This fact prompts expansion of these keywords' degree through new or recurring relations. 
Besides, new relations between old keywords $E_{o \leftrightarrow o}$ were found to be dominating which is the generic trend in inter-disciplinary research. 
Further, the growth of edges in  $E{}'_{o\leftrightarrow o}$ implies that most new hypotheses emerge across existing keywords, topics, and/or concepts. 
Conversely, sporadic nature of new hypothesis generation across new keywords can be observed through the decreasing number of edges in $E_{n\leftrightarrow n}$.
Delayed consumption of new concepts by the scientific communities can be attributed as a cause. 
In case of $E_{n\leftrightarrow o}$, we observed that a lot of edges are formed with the most central keywords within the research domain which is true for knowledge network evolution (preferential attachment). 

\subsection {Preferential Attachment and Recursive Centrality}
\setlabel{Recursive centrality}{sec:rec_centrality}
The generation of author selected keywords is governed by the inherent rules of scholarly communication which is also known as preferential attachment ~\cite{zhao2018ranking}.
In citation-based scholarly communication, few articles within a research area will draw most attention and subsequently acquire most citations.
These articles are considered
as the representatives of the corresponding research area.
Essentially, authors select keywords for their new articles either from the existing pool of keywords previously used or newly generated keywords conjoined with the existing ones.
Therefore, few representative keywords having high degree centrality can always be found in KCNs  ~\cite{choudhury2016time}. 
\begin{figure}[!h]
\begin{subfigure}{.20\textwidth}
    \includegraphics[width=1\textwidth]{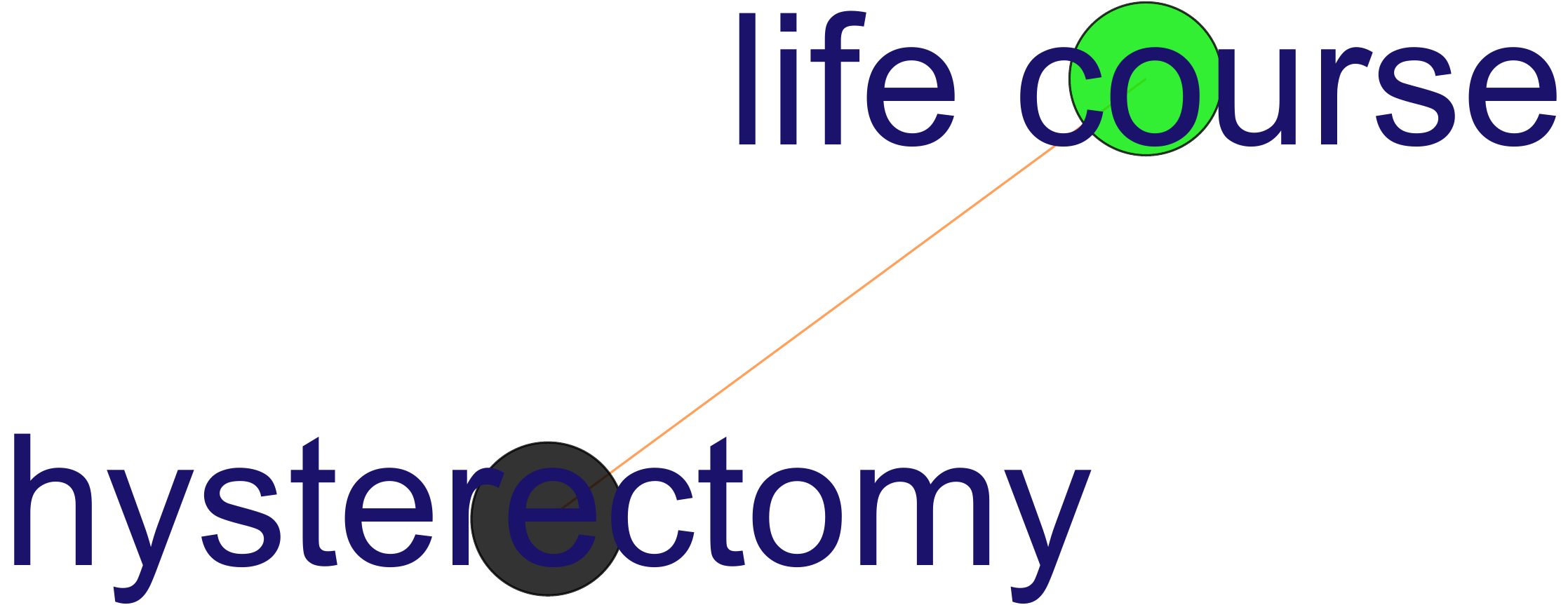}
    \caption{$G_o$: 2008 (Life course)}
    \label{fig:2011oar}
\end{subfigure}
\begin{subfigure}{.35\textwidth}
    \includegraphics[width=1\textwidth]{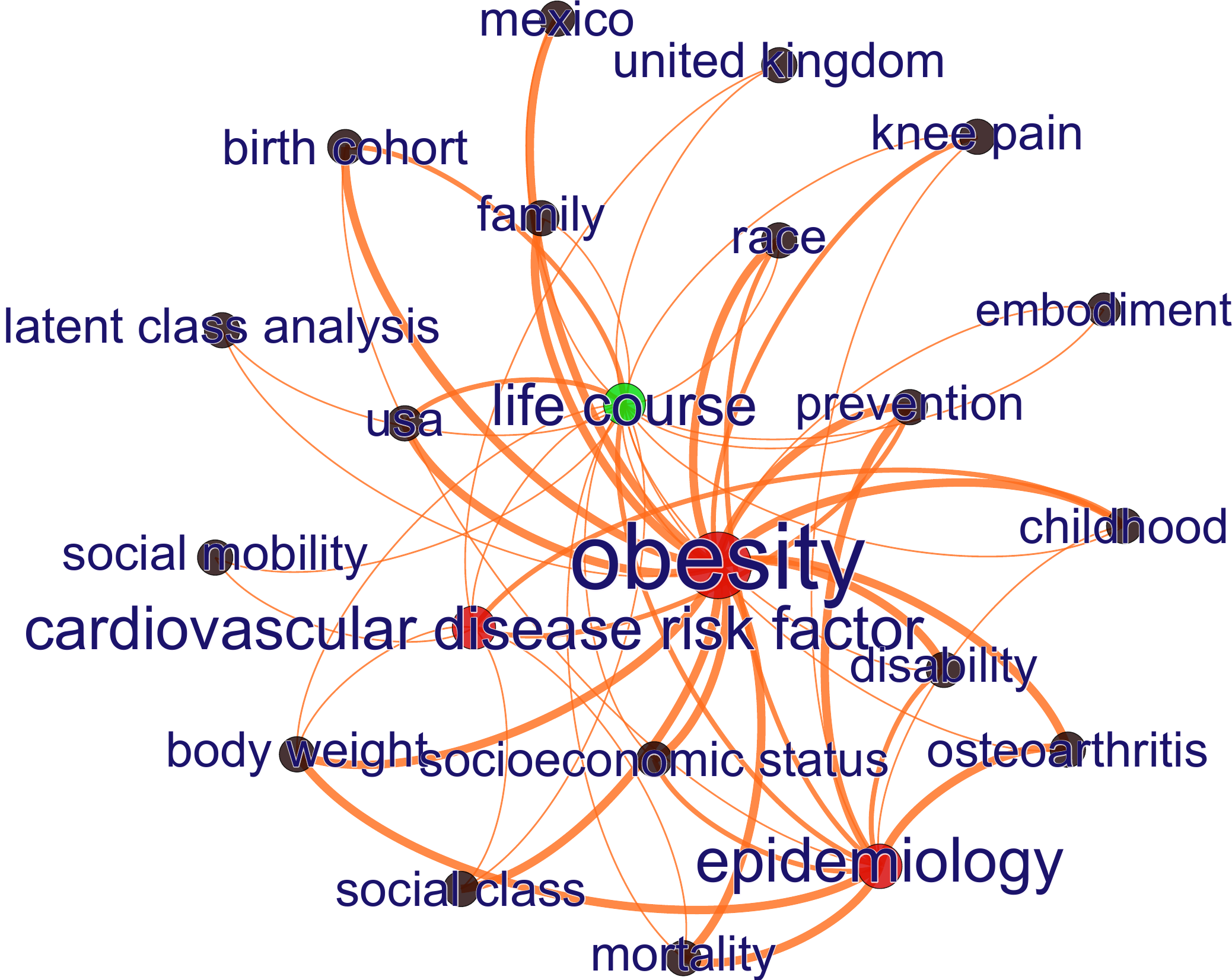}
    \caption{$G_0$: 2011 (Life course)}
    \label{fig:2011oau}
\end{subfigure}
\begin{subfigure}{.44\textwidth}
    \includegraphics[width=1\textwidth]{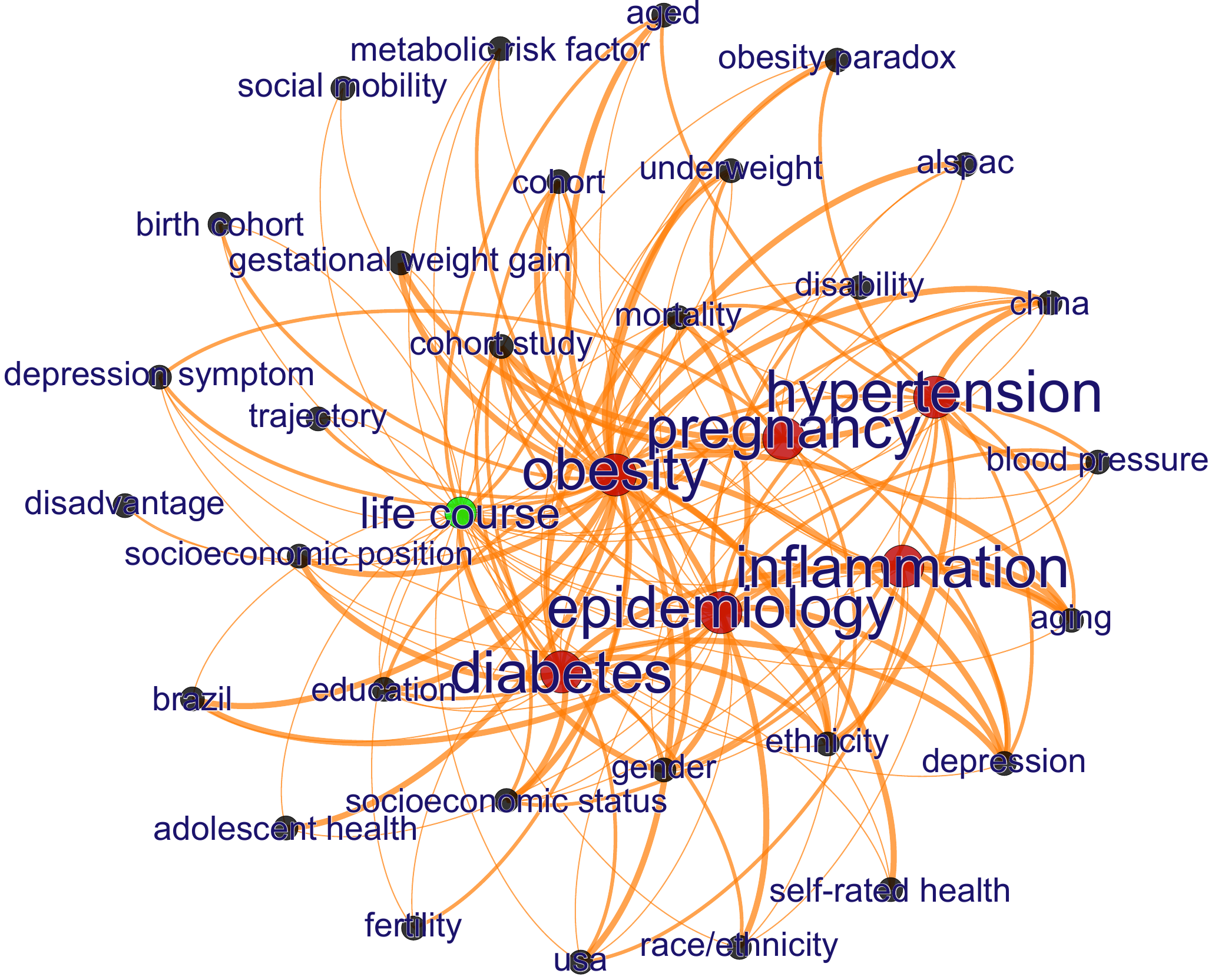}
    \caption{$G_o$: 2014 (Life Course)}
    \label{fig:2011ode}
\end{subfigure}
\begin{subfigure}{.20\textwidth}
    \includegraphics[width=1\textwidth]{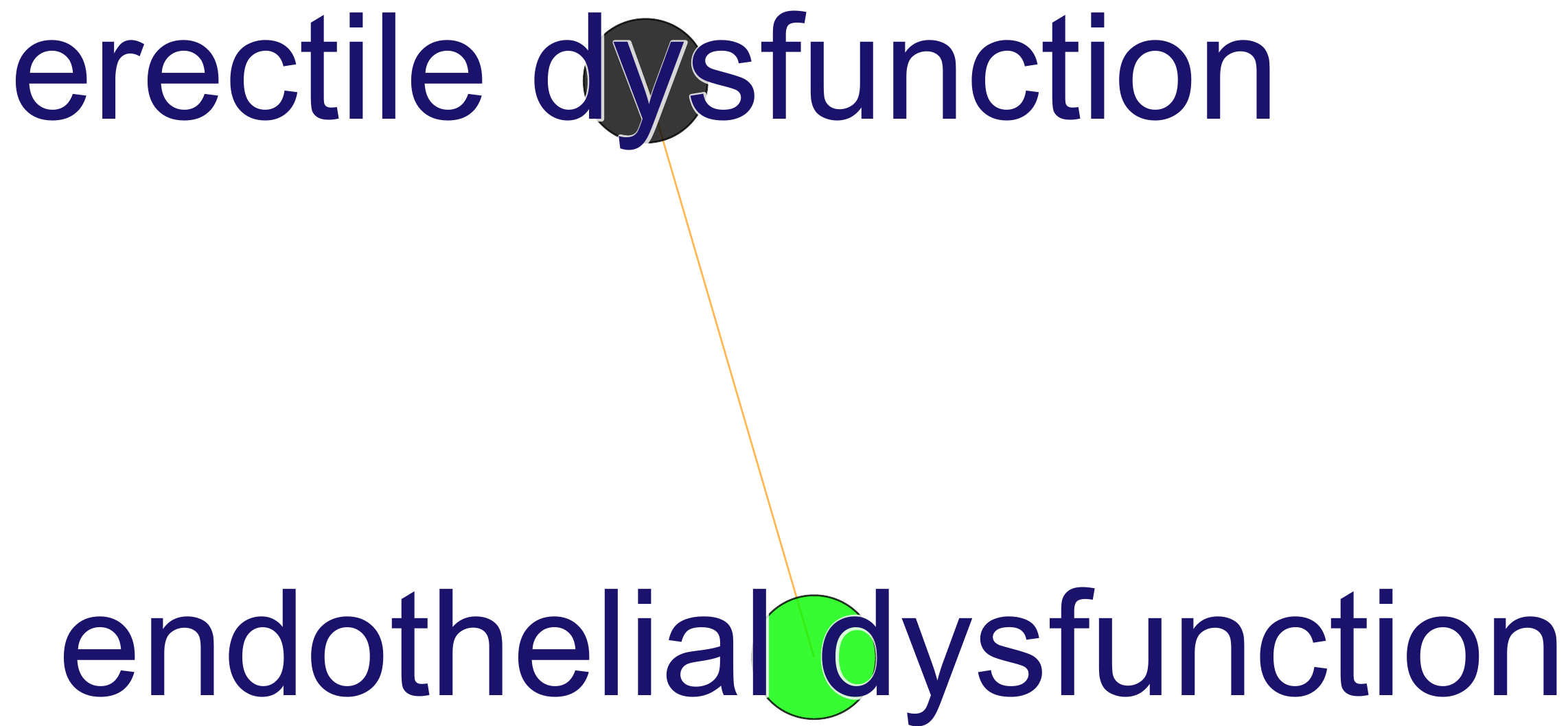}
    \caption{$G_s$: 2008 (Endothelial dysfunction)}
    \label{fig:2014sar}
\end{subfigure}
\begin{subfigure}{.35\textwidth}
    \includegraphics[width=1\textwidth]{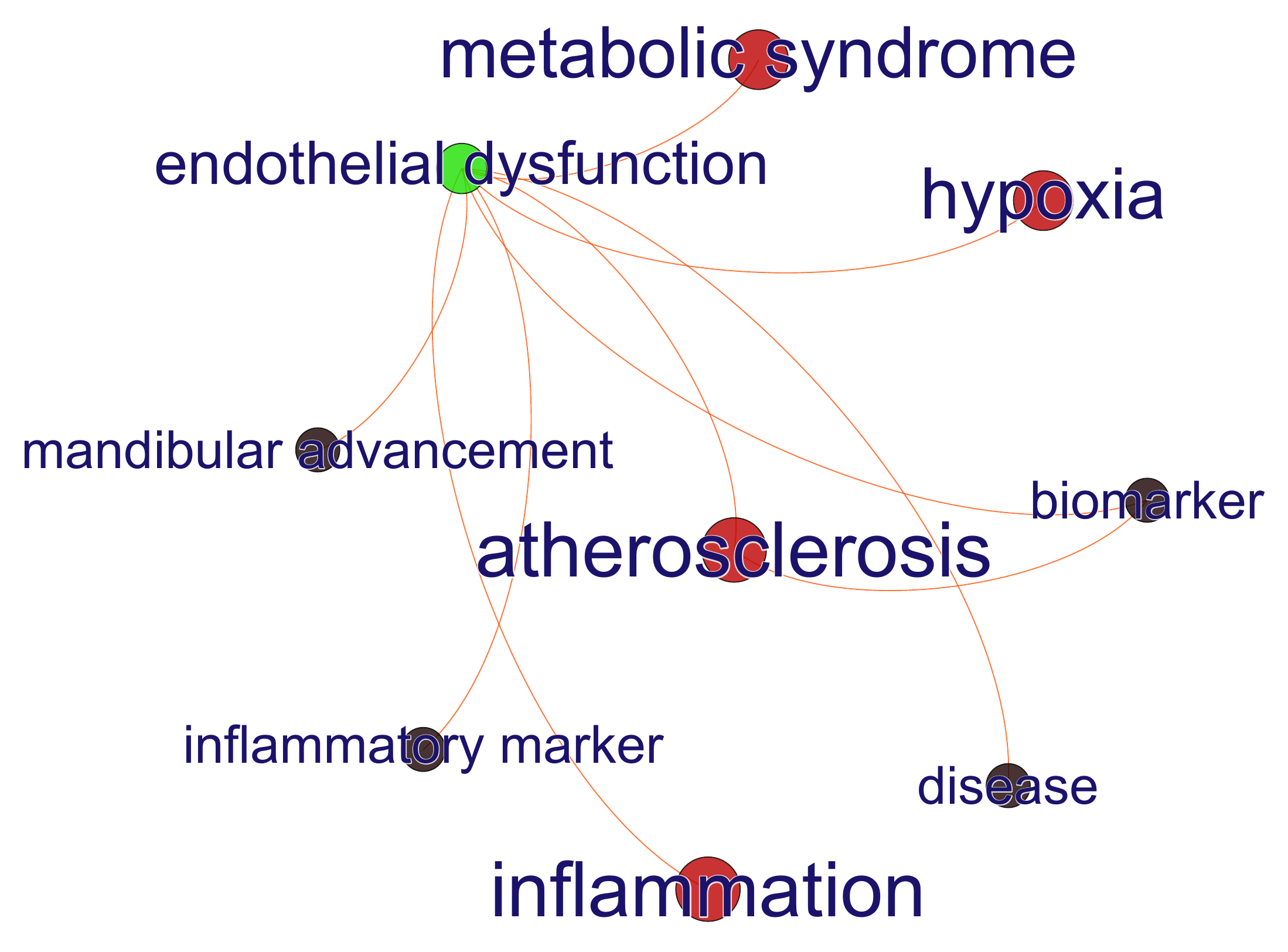}
    \caption{$G_s$: 2011 (Endothelial dysfunction)}
    \label{fig:2014sau}
\end{subfigure}
\begin{subfigure}{.44\textwidth}
    \includegraphics[width=1\textwidth]{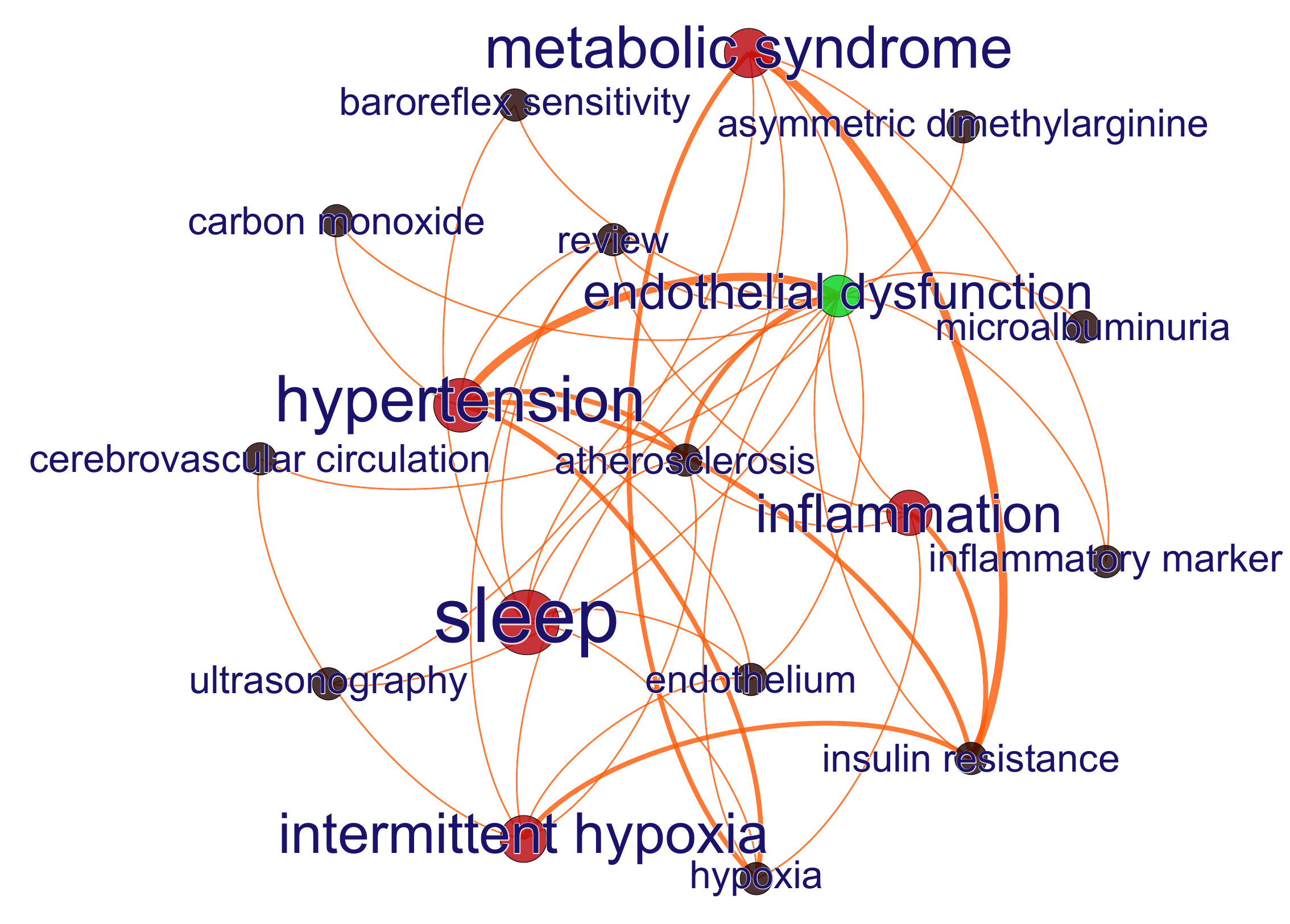}
    \caption{$G_s$: 2014 (Endothelial dysfunction)}
    \label{fig:2014sde}
\end{subfigure}
\caption {\protect\raggedright Temporal evolution of associations between new and existing keywords demonstrated by two example keywords \textbf{\textit{Life course}} and \textbf{\textit{Endothelial dysfunction}} in $G_o$ and $G_s$ respectively. 
All network snapshots are timestamped and present the evolving connection patterns of the example keywords (green coloured) with the existing keywords (red and navy coloured) of the respective research domain. The red coloured keywords are the representative keywords of the research domain and acquire more connections over time.
The size of the node represents to the corresponding keyword's degree of connections. 
The thickness of the edges represents corresponding edge weights.}
\label{fig:knowledge_evolution}
\end{figure}
This phenomenon is demonstrated in Figure ~\ref{fig:knowledge_evolution} using two keywords \textbf{\textit{Life course}} and \textbf{\textit{Endothelial dysfunction}} (green coloured) from $G_o$ (a-c) and $G_s$ (d-f) respectively in three different times. 
In these figures, the red and navy coloured keywords are the existing keywords  where the former denote the representative keywords in the respective research domain.
The size variances of the keywords represent their degree of connections in the KCN of the corresponding time.
The edge thickness represents the edge weights.
From these network snapshots, it is observable that although new keywords emerge in association with less-acquainted (relatively low profile) keywords in the corresponding research area at the beginning, however, gradually, they tend to co-appear with the representative keywords (e.g., cardiovascular disease in $G_o$ and metabolic syndrome in $G_s$).
For example, the endothelial dysfunction formed association with the keyword \textit{erectile dysfunction} in the year 2008.
In the year 2011, it form associations with the high degree keywords such as \textit{inflammation, hypoxia, metablic syndrome}.
In the year 2014, it acquire connections with more divergent high degree keywords of the domain (e.g., \textit{hypertension, sleep}).
Consequently, the rich club phenomena is also evident in dynamic KCN which can be captured by the degree centrality of the corresponding representative keywords.
\begin{table}[!h]
\begin{threeparttable}
\centering
\caption{Frequently Used Notations}
\label{notations}
\begin{tabular*}{\textwidth}{l @{\extracolsep{\fill}}ll}
\toprule
Notation    & Description \\
\midrule
${k_{gp},k_p,k_c,k_g}$        & Four keyword types: {Grandparent, Parent, Children, Guest}                          \\
$v^{au}_t,  v^{at}_t$    & Two variants of recursive centrality values for keyword $v$ at time $t$ (year) \\ & calculated from keyword-author and keyword-article bipartite relations\\
$N^{au}_{t},N^{at}_{t}$ & Sets of top $N$ central keywords in year $t$ \\ & considering the bipartite network centrality measures\\
\bottomrule
\end{tabular*}
\end{threeparttable}
\end{table}
Alongside the classical degree centrality measure, researchers attempted to define custom metrics to denote nodal importance. 
In one such study, Klimek et al.~\cite{klimek2016successful} developed a bipartite relations using term–document matrix and proposed a recursively defined document centrality measure to denote the importance of scientific documents.  
According to their assumption, a list of documents is considered central if these consume a large number of central terms those in turn also are consumed by a large number of central documents. 
Following their conceptualization of recursive document centrality measures, two recursive keyword centrality measures were developed for each keyword $v$ in each year $t$ of the training period.
Before jumping into the detail of this recursive centrality definition, Table \ref{notations} provides descriptions of some notations those will be used frequently in the sections below.

\begin{figure}[!h]
    \includegraphics[width=.98\textwidth, height=9cm]{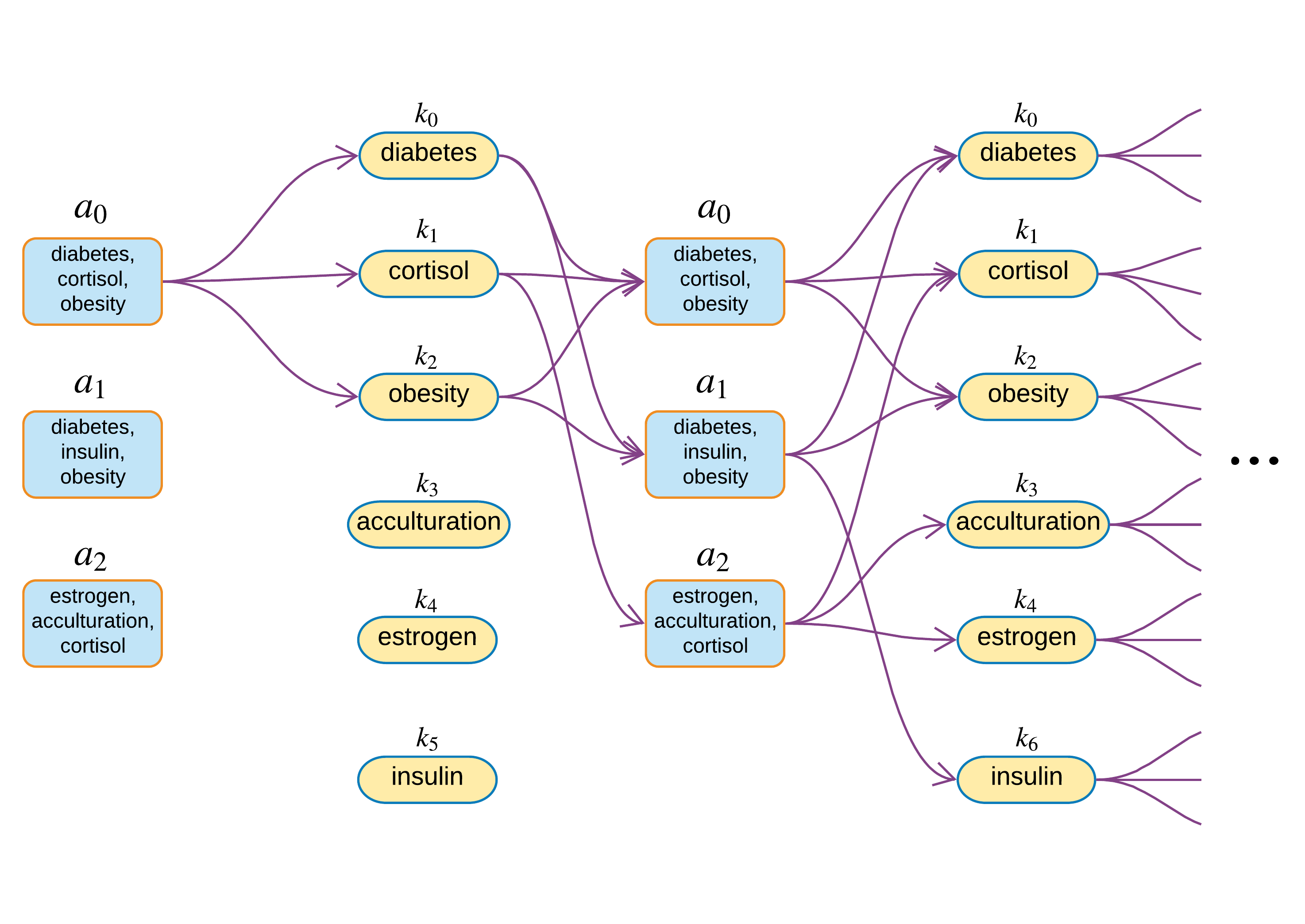}
    \caption{Computation of recursive keyword centrality measure using keyword-author relations. The computation is an iterative procedure that starts at an author, $a_0$. The first iteration counts the number of authors that used a keyword which is also used by $a_0$. Then this iteration continues until it reaches the terminating condition}
    \label{fig:origin}
\end{figure}

The first centrality measure considers the keyword-author bipartite relations whereas the second, considered the keyword-article relation.
In case of the first bipartite relations, the intuition behind a keyword's importance is defined by the number of central authors who, in turn, also use a large number of central keywords. 
After the initial textual cleaning and data processing, we filtered out  keywords which appeared in only one article.
This category denotes the specialized keywords not related to the wide majority of the scientific articles. 
With the remaining keywords, we constructed a keyword-author adjacency matrix $M_{AxK}$ for each year $t$ in the training period. 
It is worth mentioning that we considered the first name and last name of the authors and manually verified with the affiliation information to perform author name disambiguation. 
Although the objective of this manual author name disambiguation was to find all publications that belong to a given author and distinguish them from publications of other authors who share the same name, however, exploration of other methods for the same purpose is left for future studies.
The adjacency matrix $M$ is a binary matrix of size $|A| \times |K|$ where $A$ and $K$ denote the set of authors and keywords respectively in year $t$. 
An entry in matrix $M$ is $1$ if the author $a \in A$ uses the keyword $k \in K$ in his/her article.
Starting with an example author $a_i$ in year $t$, for each keyword $k_i$ used by this author, we can reach all other authors who also used $k_i$ in their article. 
Iterating this procedure twice,  we reach all the authors in two-hop distance from $a_i$. 
These authors used keyword(s) that is also used by author(s) sharing some keywords with author $a_i$. 
Figure ~\ref{fig:origin} demonstrates the iteration procedure with some example keywords.
This measure accumulates the number of different paths available among authors through their keyword usages.
Higher number represents that the corresponding keyword is pervasive across the research domain and thus more central and familiar to the key authors.
To calculate recursive keyword centrality in this way, two vectors $\psi_k$ and $\psi_a$ are defined recursively:
\begin{equation}
\psi_k(n,t)=\frac{\sum_{a}M(t)\psi_a(n-1,t)}{\psi_k(0,t)} , \  
\psi_a(n,t)=\frac{\sum_{k}M(t)\psi_k(n-1,t)}{\psi_a(0,t)} 
\end{equation}
with $n$ represents the number of iterations as portrayed in Figure ~\ref{fig:origin}. $\psi_k(0,t)$  and $\psi_a(0,t)$ denote the initial conditions  as $\psi_k(0,t)=\sum_{a\in A} {M}(t)$ and $\psi_a(0,t)=\sum_{k \in K} {M}(t)$. 
The values of $\psi_k$ can have different interpretations.
The initial condition $\psi_k(0,t)$ denotes the degree centrality of a keyword.
For different values of $n=1, 2, 3 ...$, values of $\psi_k$ assign weights to individual keywords considering the number of authors consuming them in their articles. 
Therefore, high values of $\psi_k$ corresponds to the keywords selected by a large number of authors those used large number of such keywords.
Conversely, low values represent the keywords` specificity and relevance to limited number of research issues. 
Similar observations are also true in case of $psi_a$.
The number of iteration converged  and found stable at $n=20$.
We also attempted higher value of $n$ however. found similar recursive centrality values returned by the algorithm.
Thus, for each year $t$ in the training period, we calculated a vector of z-score normalized recursive keyword centrality values $C^{au}_t$ by following the algorithm, from the keyword-author bipartite relations.
Likewise, another recursive keyword centrality measure $C^{at}_t$ was constructed from the keyword-article bipartite relations in each year $t$ during the training interval. 
In addition to these recursive centrality measures, to compare the results with a traditional centrality measure, we also calculated degree centrality values of keywords $C^d_t$, extracted from the unipartite keyword-keyword relations in KCN for each year $t$.
Future studies can calculate other centrality values to compare the results further.

\subsection {Genealogical Typology}
\setlabel{Genealogical Typology}{sec:genealogy_typology}

As mentioned earlier, due to prevailing preferential attachment in KCN, there exists a set of most central keywords which exercise greater influence over the structure of the network. 
It was observed that other keywords generally tend to form relations by co-appearing with these rich keywords.
To predict the future co-evolution of keywords, this fact is needed to take into consideration.
In this study, we capitalize the idea of preferential attachment, but in a different way.
\begin{figure}[!h]
    \centering
    \includegraphics[width=.9\textwidth, height=8.5cm]{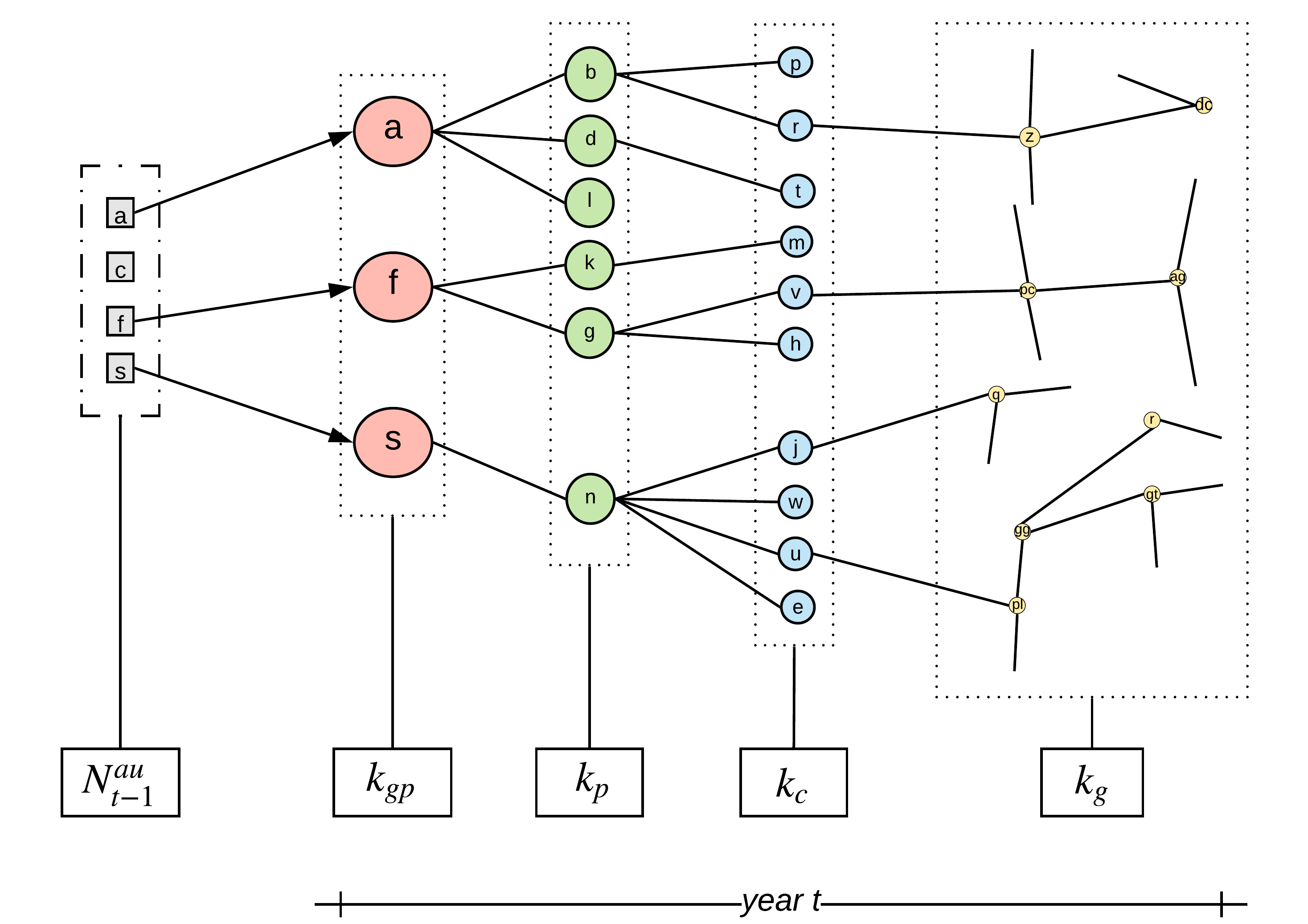}
    \caption{Genealogical communities of keywords by considering a set of keywords with high recursive centrality values (i.e., $N^{au}_{t-1}$) computed from keyword-author bipartite relationships. Keywords with high centrality values at timestamp $t-1$ are designated as grandparents ($k_{gp}$ at timestamp $t$. Direct neighbours of these grandparents belong to parents ($k_p$) community, and two-hop neighbours are designated as children ($k_c$).Rest of the keywords are designated as guests ($k_g$).} 
    \label{fig:heredity_build}
\end{figure}

We defined a keyword's genealogical community membership in each year that denotes it's current ancestral relation with the central keywords of the previous year. 
The underlying objective here was to capture the evolution of the a keyword's lineage to its ancestry. 
Genealogy is the study of family tree and a genealogy graph is used to portray the complex evolving relation originating from the ancestors. 
In the same way, we defined a keyword's family relationship in a particular year $t$ based on it's relation with the central keywords from previous year $t-1$. 
The centrality of keywords was measured by considering two recursive centrality measures defined above and one degree centrality.
This approach would label keywords according to the type of their ancestral relationships (e.g., grandparents, parents, child or guest) and help us understanding the impact of these relationship types in their co-occurrences.
The construction of the genealogical communities is described below:

In our dynamic KCN, we defined four types of communities for keywords, namely, grandparents $(k_{gp})$, parents $(k_p)$, children $(k_c)$ and guests $(k_g)$.
Firstly, we defined the grandparents.
If we consider identifying community membership of keywords in year $t$, then grandparents $k_{gp}$ keywords are the top-$N$ central keywords from year $t-1$. 
Here, the word `central' denotes higher values of a chosen centrality measure from the three aforementioned centrality measures (i.e., two recursive and one degree centrality). 
In other words, these are the most frequent and influential keywords both in regards to the contents of metadata information (e.g.,author and article) and in forming relations with other keywords. 
We experimented with the number of grandparents ranging from $10-200$ by considering two recursive centrality measures.
Interestingly, we found that having $N\geqslant20$  grandparents in each year does not change the prediction performance (described later) significantly in both datasets.
Therefore, we designated the top $20$ keywords as grandparents by considering the chosen centrality measures. 
In this way, we got three sets of top-$20$ central keywords  based on our three centrality measures (i.e., $C^{au}_{t-1}$, $C^{at}_{t-1}$, and $C^{d}_{t-1}$). 
Secondly, we define the parents, children and guests.
Keywords having direct relations (i.e., an edge is observed between them) with $k_{gp}$ are labeled as parents $(k_p)$ in the year $t$. 
Subsequently, children $(k_c)$ are the direct neighboring nodes of $k_p$ keywords but not $K_{gp}$. 
Alternatively, $k_p$ keywords are considered as common neighbours between $k_{gp}$ and $k_c$. 
Once these three communities of keywords (i.e. $k_{gp}, k_p, k_c$) are labeled, rest of the are designated as guests $(k_g)$. 
These guests keywords had maximum distance from the grandparents.
In this way, we defined the genealogical communities of all keywords appearing in a particular year. 
In our experimental setup, we use top-$20$ keywords from year $(t=0)$ (i.e., 2007) for the first training year $(t=1)$ (i.e., 2008), though this year $(t=0)$ is not included within our training period. 
This fact was mentioned in the earlier section \ref{sec:data_acquisition}.
\begin{figure}[!h]
\begin{subfigure}{.33\textwidth}
    \includegraphics[width=1\textwidth]{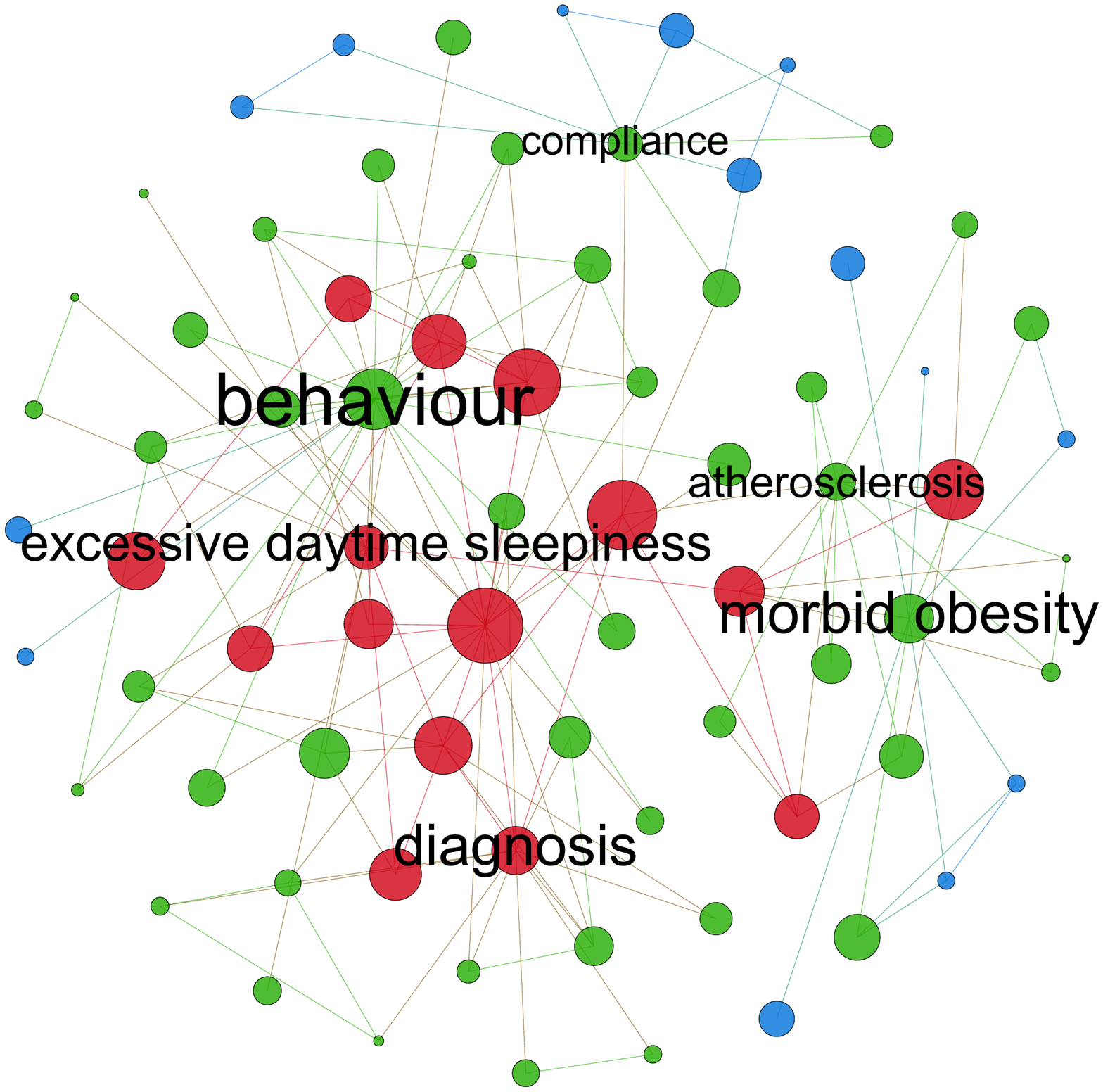}
    \caption{$G_s$: 2011 (keyword-article)}
    \label{fig:2011oar}
\end{subfigure}
\begin{subfigure}{.33\textwidth}
    \includegraphics[width=1\textwidth]{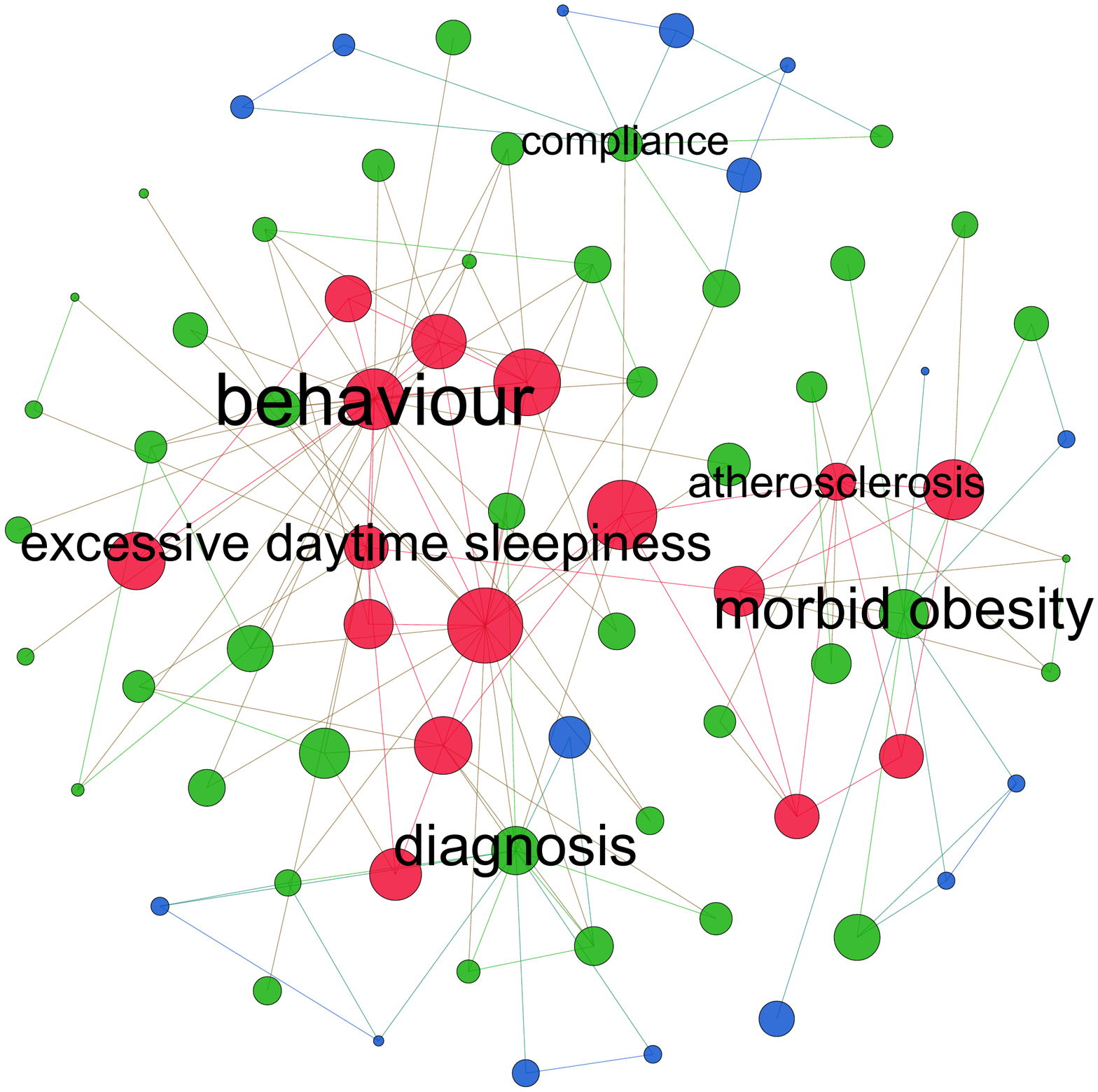}
    \caption{$G_s$: 2011 (keyword-author)}
    \label{fig:2011oau}
\end{subfigure}
\begin{subfigure}{.33\textwidth}
    \includegraphics[width=1\textwidth]{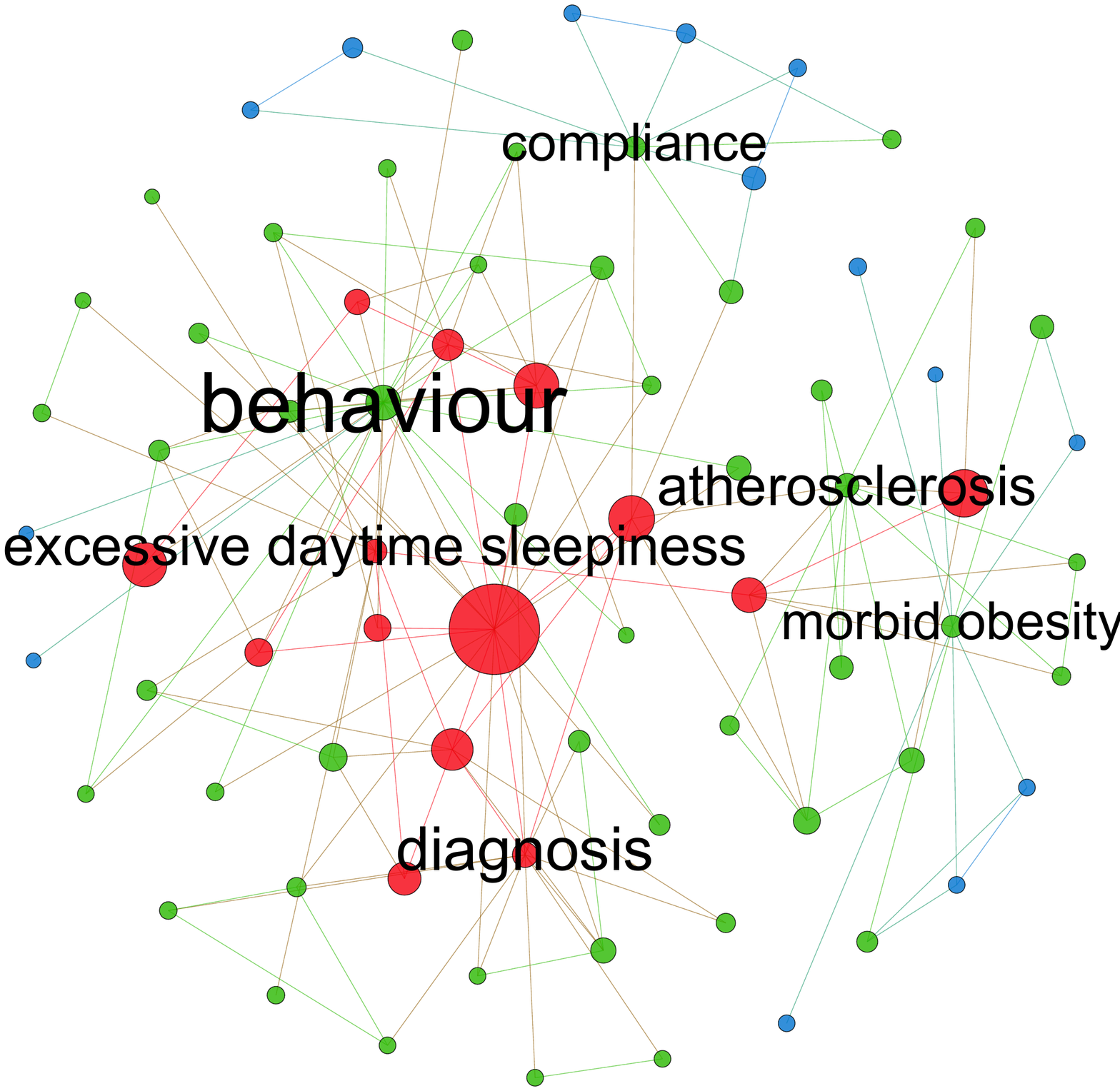}
    \caption{$G_s$: 2011 (degree)}
    \label{fig:2011ode}
\end{subfigure}
\begin{subfigure}{.33\textwidth}
    \includegraphics[width=1\textwidth]{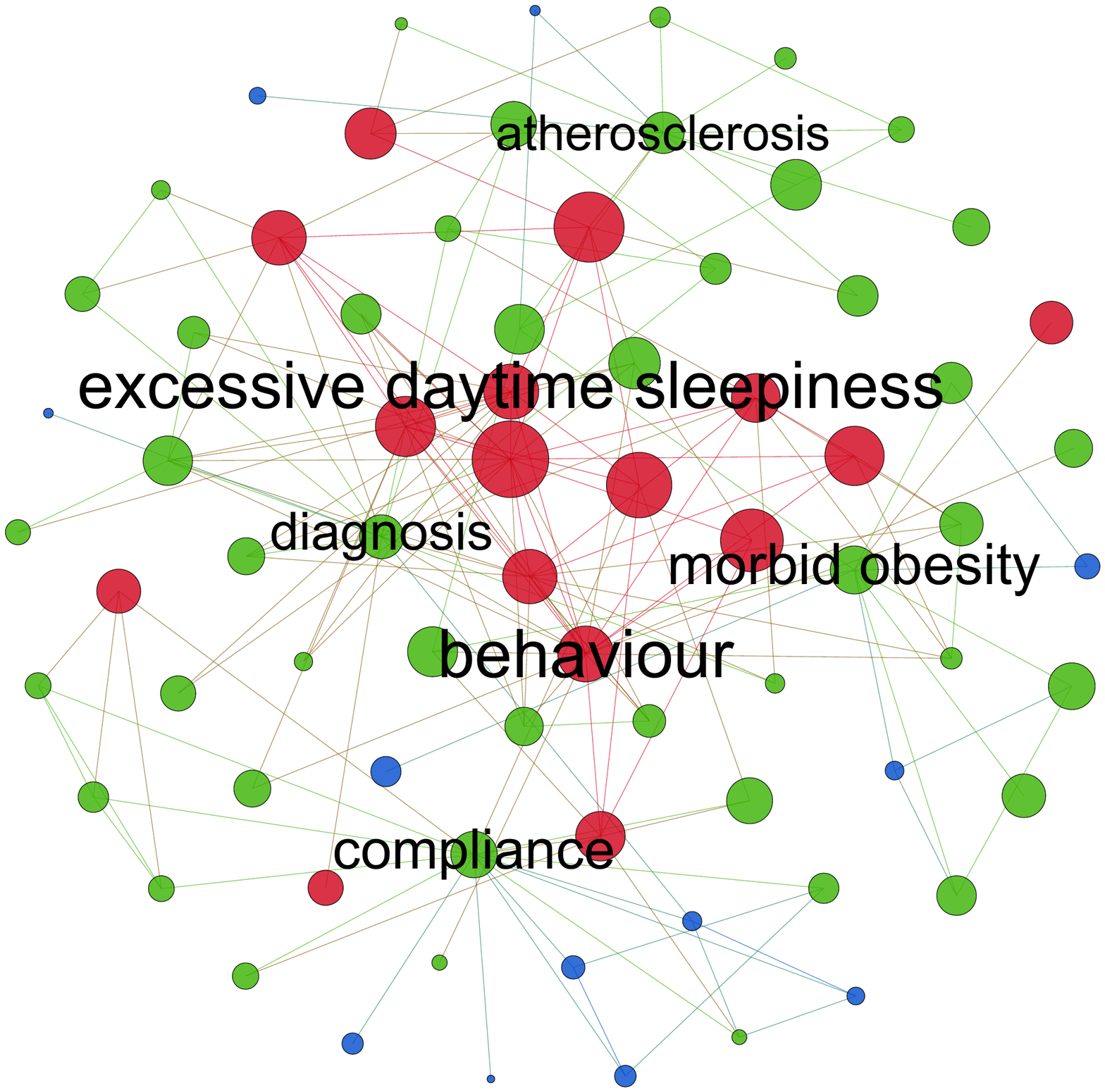}
    \caption{$G_s$: 2014 (keyword-article)}
    \label{fig:2014sar}
\end{subfigure}
\begin{subfigure}{.33\textwidth}
    \includegraphics[width=1\textwidth]{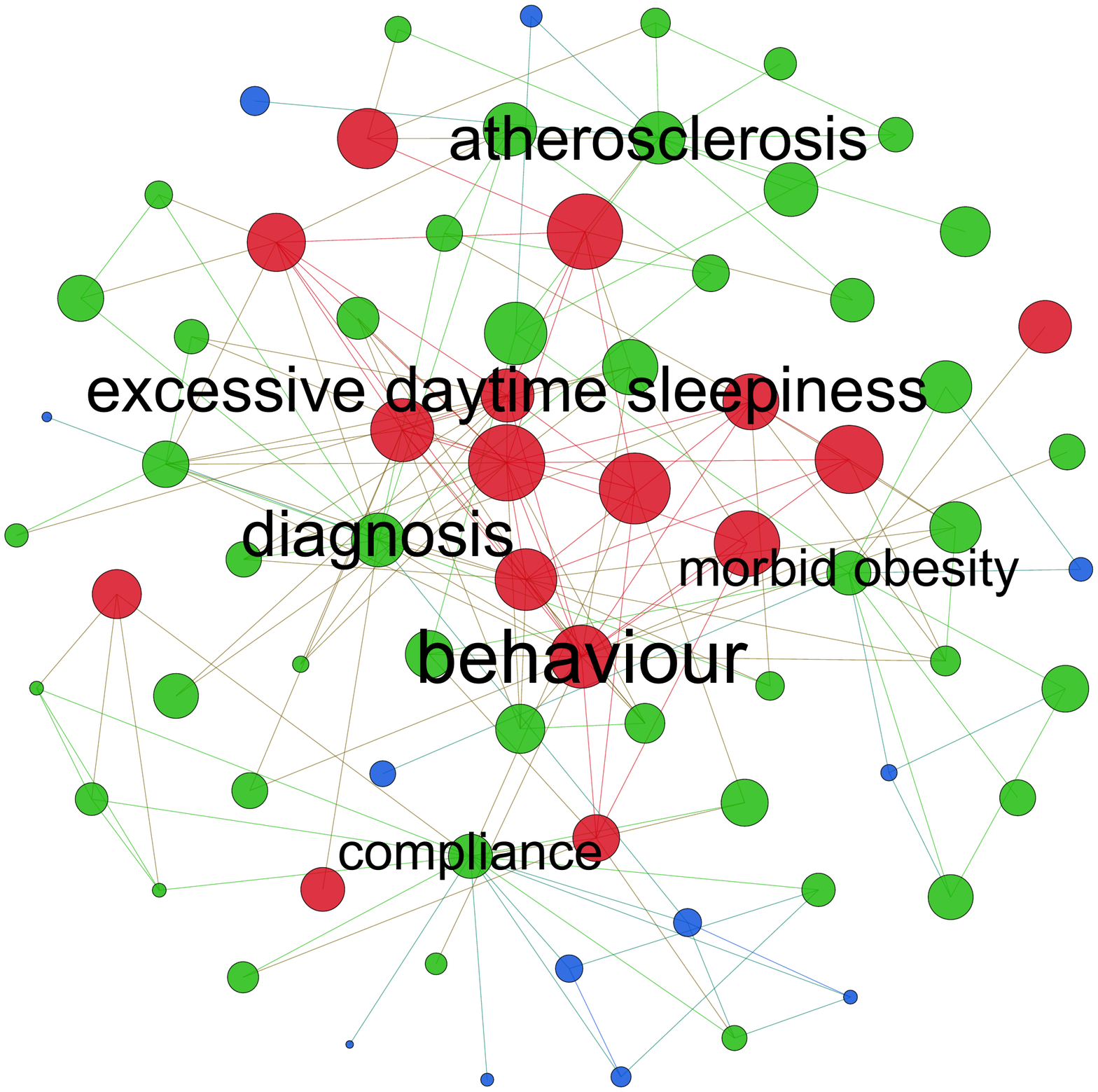}
    \caption{$G_s$: 2014 (keyword-author)}
    \label{fig:2014sau}
\end{subfigure}
\begin{subfigure}{.33\textwidth}
    \includegraphics[width=1\textwidth]{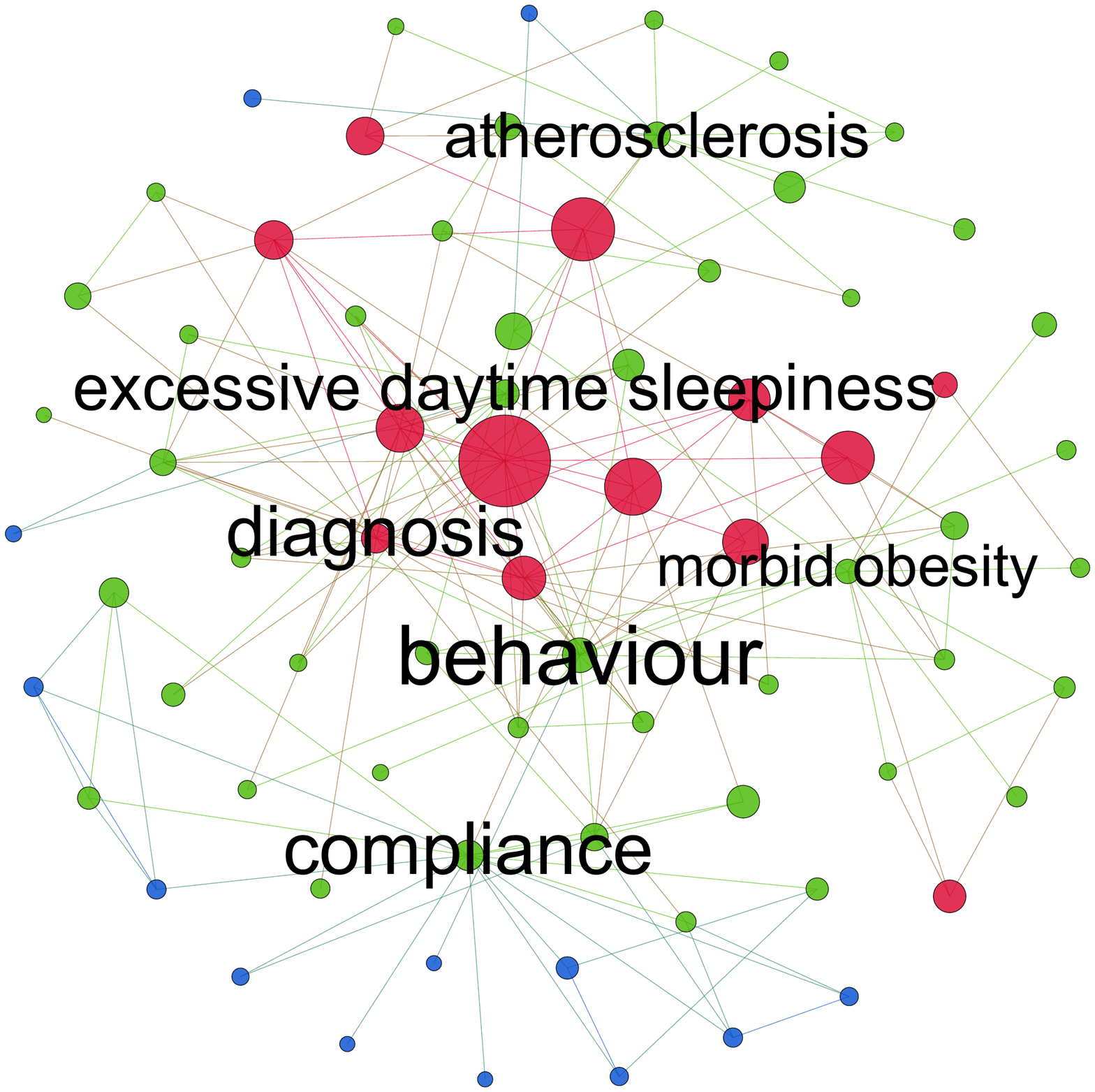}
    \caption{$G_s$: 2014 (degree)}
    \label{fig:2014sde}
\end{subfigure}
\caption {\protect\raggedright Temporal variations of genealogical traits demonstrated by keywords depending on the centrality measures used in $G_o$ dataset. 
All network snapshots are timestamped including the centrality measure used to identify the genealogical traits.
The color codes represent the genealogical communities of keywords: red (grandparent), green (parent) and blue (child). 
The size of the node represents to the keyword's corresponding centrality measures. }
\label{fig:h_variation}
\end{figure}
For the purpose of constructing link prediction features, we assigned each keyword a score according to its genealogical community membership. 
The successive order of keyword's genealogical community score is: $score(k_{gp}) >> score(k_p)> score(k_c)> score(k_g)$.

Being most influential in the previous year, the grandparents (i.e., $k_{gp}$) were assigned the highest score in a year. 
Scores for the other three keyword types (i.e., $k_p$,$k_c$,$k_g$) were defined according to their distance from the grandparents $k_{gp}$. 
Being the most distant ones, $k_g$ (guests) are assigned with lowest community score. 
The process of assigning a keyword's genealogical community scores is depicted in figure \ref{fig:heredity_build}.
The relative size of each keyword in this figure represents the weight of the score (i.e. greater size for higher value). 
In this figure, $N^{au}_{t-1}$ denotes the set of ton-$N$ central keywords by considering recursive centrality values $v^{au}_t$ extracted from keyword-author relations.

Considering three sets of centrality measures of keywords (i.e., $v^{au}_t,  v^{at}_t$  and $v^{d}_t$), a keyword belonging to one genealogical community (\eg, grandparent) in one set, might belong to different community(\eg, parent) in another.
This effectively means that if a keyword becomes grandparent by considering the $v^{au}_t$, it may not necessarily be true if the other centrality measures (i.e., $v^{at}_t$ or degree centrality) were considered.
Likewise, considering the same centrality measure, a keyword can belong to different genealogical community in different year(s).
In Figure \ref{fig:h_variation}, a comparative representation of these variations is presented  for some keywords in $G_s$ datasets.
In this figure, top two rows present network snapshots in $G_s$ considering six keywords (i.e., \textit{\textbf{atherosclerosis, morbid obesity, diagnosis, excessive daytime sleepiness, behavior, and compliance}}).
(It is evident that there exists some commonality of keywords in both datasets since some of these keywords resemble to those in the $G_o$ dataset.
Since the commonality is out of the scope of this study, we leave it for future studies where domain experts can contribute towards literature based discovery related to these two domains together.)
The first row in this figure presents the network snapshots in the year 2011 and the second presents snapshots in the year 2014.
In each row, the left snapshot presents the genealogical typologies (i.e., communities) of keywords identified by considering the recursive centrality (i.e., $v^{at}_t$) from keyword-article bipartite relations.
The snapshots in the middle column present the genealogical typologies of keywords identified by considering the recursive centrality  (.i.e., $v^{au}_t$) from keyword-author bipartite relations.
The right column presents the similar by considering the keyword's degree centrality.
Color codes represent different types of the keywords: grandparent (red), parent (green), and child (blue).  
The sizes of the nodes and labels denote corresponding centrality values.
It is observable that keywords can belong to different communities in different years by considering different centrality measures used in this study.
Consider the genealogical communities of the keyword \textit{diagnosis}.
In the year 2011, it was designated as `grandparent' by considering  $v^{at}_t$, whereas considering the same centrality measure, it belonged to `parent' community in 2014.
Surprisingly, the later is also true (`parent') when recursive centrality measure $v^{au}_t$ was considered in the year 2011.
Similar observations are evident in case of the keyword \textit{behaviour}
There are some domain-specific keywords (\eg   \textit{sleep}, \textit{apnea},\textit{ obesity}) which were found as `grandparents' in all centrality variants since they are the representative keywords of the respective domain. 
\begin{table}[!h]
\begin{threeparttable}
\centering
\caption{Number of nodes (keywords) according to different genealogical typologies (i.e., communities) defined in this study and edges (keyword pairs) between different typologies of keywords. $GP$ represents the grandparents keywords, $P$ represents the parents, $C$ represents the children and $G$ represents the guests keywords. Similarly, $E_{P}$ denotes the edges between parent keywords, and $E{GP \leftrightarrow P}$ denotes the edges between grandparent and parent keywords.}
\label{table:gr_type}
\begin{tabular}{lrrrrrrrrrrrrrrr}
\toprule

Year &   $P$ &  $C$  &  $G$ &    $E_{GP\leftrightarrow P}$ &  $E_{GP \leftrightarrow C}$ & $ E_{GP \leftrightarrow G}$ & $E_P$ &  $E_{P \leftrightarrow C}$ & $E_{P \leftrightarrow G}$ & $E_C$ &  $E_{C \leftrightarrow G}$ & $E_G$\\
\midrule
& & & & & &$G_s$\\
 \midrule
2008 &      197 &   95 &  147 &         66 &      27 &      52 &   91 &     70 &     86 &   15 &     46 &   26 \\
2009 &     179 &   84 &  187 &           65 &      32 &      56 &   89 &     62 &     95 &   13 &     30 &   32 \\
2010 &     189 &   98 &  189 &            70 &      29 &      58 &   97 &     48 &    101 &    9 &     34 &   26 \\
2011 &     225 &  116 &  175 &           81 &      29 &      52 &  112 &     64 &    115 &   13 &     41 &   25 \\
2012 &    243 &  127 &  194 &          103 &      51 &      49 &  151 &     94 &    104 &   19 &     49 &   23 \\
2013 &    273 &  167 &  174 &          99 &      45 &      52 &  156 &    121 &    114 &   23 &     49 &   24 \\
2014 &    235 &  156 &  172 &       100 &      53 &      35 &  141 &    140 &     66 &   31 &     64 &   27 \\
\midrule
& & & & & &$G_o$\\
\midrule
2008 &    1765 &  430 &  141 &          679 &     185 &      67 &  3729 &   1083 &    283 &   72 &     38 &    3 \\
2009 &     1614 &  360 &   98 &          664 &     157 &      32 &  3626 &    968 &    242 &   63 &     44 &    0 \\
2010 &     1752 &  409 &  117 &           556 &     153 &      43 &  4072 &   1148 &    302 &   69 &     25 &    5 \\
2011 &     1963 &  334 &   91 &          665 &     125 &      27 &  4628 &    925 &    258 &   37 &     34 &   10 \\
2012 &     2109 &  368 &   82 &           757 &     131 &      26 &  5723 &   1024 &    190 &   48 &     12 &    1 \\
2013 &     2117 &  370 &   75 &         726 &     148 &      41 &  5582 &   1375 &    211 &   73 &     30 &    1 \\
2014 &    2252 &  793 &  155 &       869 &     323 &      51 &  6903 &   2200 &    310 &  193 &     54 &    4 \\
\bottomrule
\end{tabular}
\end{threeparttable}
\end{table}

\subsection {Genealogical Community and KCN Evolution}
\setlabel{Keyword's Community and KCN Evolution}{subsec:type_evolution}

In Table \ref{table:gr_type}, basic statistics of nodes (keywords) and different types of edges (keyword co-occurrences) formed between keywords belonging to different genealogical communities are presented. 
In this table, we ignored the grandparent keywords since for each year it was the top-20 keywords from the previous year.
Consequently, edges among the grandparent keywords were also ignored since they were trivial in comparison to the other types of edges.
In both datasets, we found that edges among keywords belonging to the parent community dominate in numbers.
This fact is correlated with the increase of number of edges between grandparents and parents over time.
Considering the sizes (i.e., number of articles and keywords) of both datasets, $G_s$ harbours more guest keywords and edges.
However, the ratio of edges between children and guest keywords are similar in both datasets.
The children and guest keywords tend to form edges with parents more than the grandparents.
This chronology of descendants not only helped us understand the temporal trends of topics and research hypotheses developments but also attribute specificity in preferential attachment.
For example, in absence of such typologies, we would consider grandparents and parents keywords not only in the same category and but also the richest since these would acquire most of the emerging edges.
From this table, we can also observe contrasting phenomenon which signifies the fact that the richest does not always remain so as they grow oldest.
Such classification on genealogical traits will allow us to comprehend the evolutionary growth of KCNs better and thus help in science mapping.

\begin{figure}[t]
    \centering
    \includegraphics[width=1\textwidth]{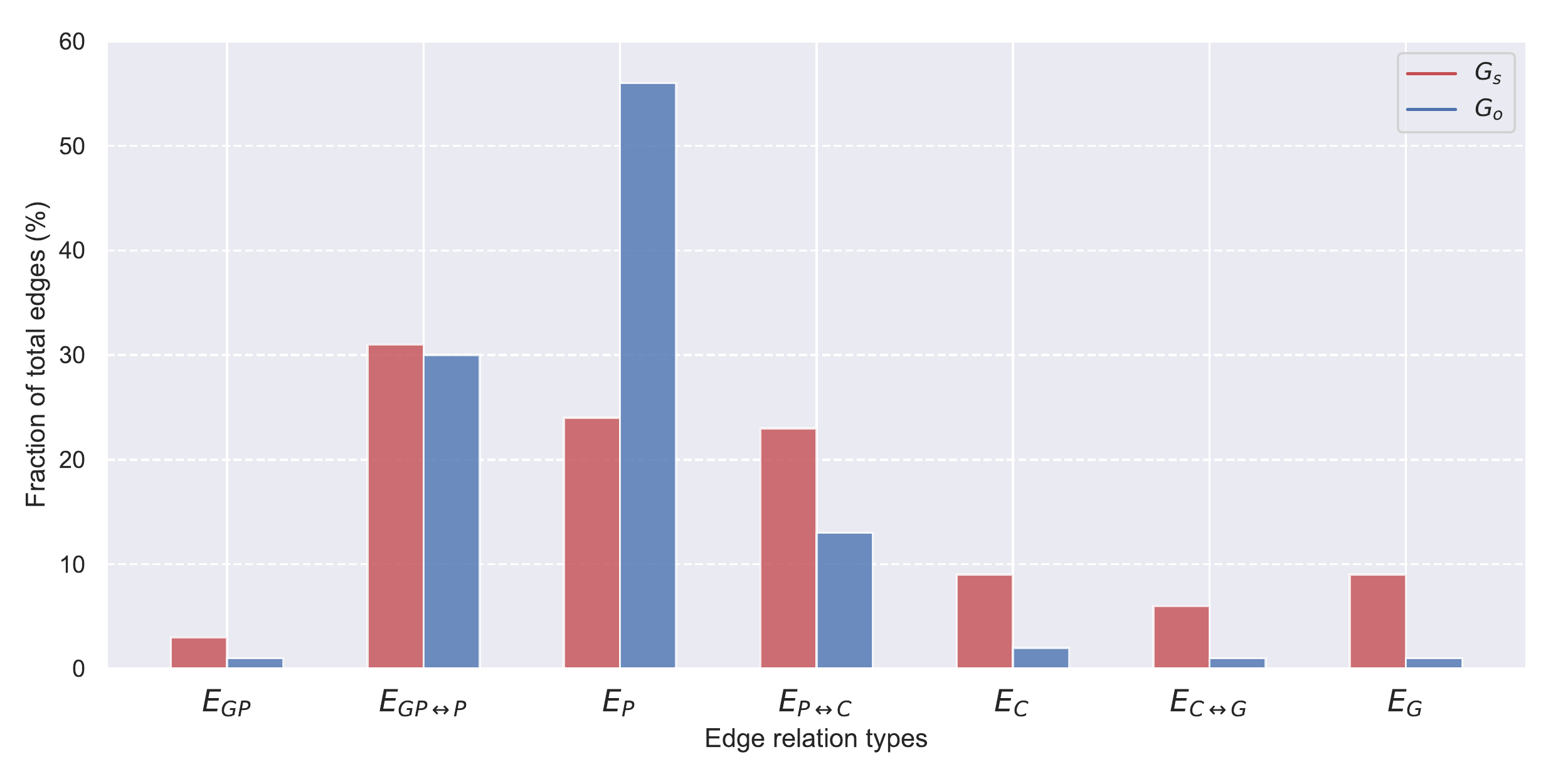}
    \caption{Percentage of different types edges among keywords from different genealogical communities with respect to the total number of edges in the test year (i.e., 2015)}
    \label{fig:test_relation}
\end{figure}

In Figure ~\ref{fig:test_relation}, we observe how this ancestral relationships helped us to classify the edge types in the test year (i.e., 2015).
In this figure, we present the percentage of different types of edges, as presented in Table \ref{table:gr_type}, with respect to the total number of edges.
It is evident from this figure that the keywords belonging to the parent community play crucial roles in emerging relations than any other keyword types.
In both datasets, the parent keywords dominate in attracting both descendants and antecedents to form emerging relations.
\section{Research Methodology}
\setlabel{Research Methodology }{sec:research_method}
We frame the  future LBD prediction problem
as a learning-based link prediction task.
Predicting co-evolution of author selected keywords or identifying implicit relationships between them can be mapped to the dynamic supervised link prediction model from network science.
In this section, we describe our research methodology to construct such a predictive model that incorporates both nodal and edge features, and a recurrent neural network for both feature forecasting and classification.

\subsection {Dynamic Supervised Link Prediction}
\setlabel{Link prediction problem formulation}{subsec:link_prob}
For the purpose of link prediction, the total duration in each dataset was split into two non-overlapping intervals, $T$ and $T+1$, known as the training and test phase. 
The primary objective of dynamic supervised link prediction framework is to analyse the temporal topological structure and nodes` attributes in the network of training phase $G_T(V_T,E_T)$ to predict the emergence of future edges in the network of test phase $G_{T+1}(V_T,E_{T+1})$. 
In dynamic network perspective, $G_T$  is sampled using an aggregation granularity to generate a time series of network snapshots. 
Due to the data collection granularity this study used yearly window size to sample the network $G_T$ and generate time series of network snapshots. 
It is impractical to predict the edges between nodes those are absent in the training network. 
Therefore, $V_T$ is the set of nodes appeared in both phases.
Then, we prepared our classification dataset consists of the instances of non-connected keyword-pairs in the training phase. 
Each instance is labeled either positive or negative based on it's presence  as a true link in the test network.
Supervised method for link prediction problems needs to predict the emerging edges by successfully discriminating the positive and negatively labelled keyword-pairs within the classification dataset. 
Hence, supervised link prediction is considered as a binary classification task by learning positive and negative instances with the help of interesting features describing each instance.
Classification model requires effective features for the training purposes. 
To perform this task, we constructed novel features (described later) by incorporating temporal information, network structure, and genealogical community memberships of author selected keywords.
The feature values for both positive and negative instances were computed by considering each network snapshot at timestamps ${t_0, t_1, t_2, ... t_n}\in T$.
Here, each timestamp represents a year in the training phase $T$ (i.e., 2008-2014) .
A recurrent neural network (described below) was also employed to forecast the feature values for each instance in the test phase.
These forecasted feature values were input into a classifier for the classification purpose.

To measure performance of the classification, there are mainly two categories of evaluation metrics in a supervised learning problem: (i) Fixed threshold metrics like accuracy, precision, recall (ii) k-equivalents and Threshold curves like precision-recall (P–R) curve and the area under the ROC curve (AUC) ~\cite{yang}.
The threshold curves like ROC curve and P-R curve are two dimensional curves. 
In ROC curve, true positive rates are plotted on the Y-axis and false positive rates are plotted on the X-axis. 
In P-R curve, Y-axis is for precision and X-axis is for recall.
P-R curve provides better measurement in case of class distribution with large skewness.
It also performs better in providing more insight regarding the exposure of class differences. 
\subsection {Feature Engineering}
\setlabel{Feature Engineering}{subsec:feature_eng}
To support supervised link prediction task, we constructed different features by taking advantages of different recursive centrality measures and genealogical communities extracted from temporal KCN.   
These features assessed the influence of keyword in regards to authors and articles and leveraged the temporal significance of evolving networks.
Before generating features for keyword pairs (edge-level), we first identified different characteristics of individual keyword's importance. 
The rationale behind this is to contemplate different levels of significance each keyword carries in regards to both network importance and ancestral lineages.
We also considered accumulated citation score over time for each keyword to construct citation-based features.
Since, these features only pertain to one keyword, some aggregation functions are needed to combine/aggregate the individual feature value of the corresponding keywords in a keyword pair.
To illustrate further, consider the fact that if either(or both) of the keywords are prolific or belong to the same community, it is more likely that they will co-occur.
Before aggregation, the individual measure denotes the proliferation rate of individual keyword,  its community membership, or temporal activeness (number of co-occurrences).
Aggregation of these individual features yields the aggregated features which are meaningful for the pair of keywords in dynamic link prediction.
Therefore, keyword-specific features were then aggregated to develop features for each keyword pairs.
In this study's context, we assume that the higher the aggregated feature values for a keyword pair, the more likely that those two keywords will co-occur.
In the following section, we describe the features used in the classification task.
\subsubsection{Keyword centrality}
\setlabel{Keyword centrality}{subsec:key_cen}
For each keyword, we considered three centrality measures as node-level features. 
Two recursive centrality measures were based on the keyword-author ($v^{au}_t$) and keyword-article ($v^{at}_t$) bipartite relations as defined in section \ref{sec:rec_centrality}.
The final centrality measure was the corresponding keyword's degree centrality ($v^{d}_t$).
\subsubsection {Temporal community importance}
\setlabel{Temporal community importance}{subsec:heredity}	

This is the first edge-level aggregated  feature we computed for each non-connected keyword pair for the link prediction purpose.
The objective of this feature was to capitalize both the origin information (genealogical community) and current activeness (centrality measures) of both keywords in a keyword pair.  
Chenhao Tan ~\cite{genealogy} investigated how the origin of a community connects to its future growth and demonstrated that the history of the early members allows us to understand the emergence of communities (i.e., tracing the origin of a community).
In regards to the future community growth, the author believed that the emerging process of a community is analogous to complex contagion that requires dense connections between early adopters.
In our case the early adopters are the members of the grandparent and parent communities.
Therefore, we first multiply individual keyword's community score  with its corresponding centrality value  for each year in the training period.
The former denotes the keyword's relation with previous year's central keyword (i.e., keyword's family lineage) and the later denotes its importance which was originally used to determine its community membership.
This will augment the community importance scores of grandparent and parent keywords more than children and guests, and subsequently determine the individual temporal community importance score of a keyword in each training year.
Finally, for each keyword pair, the temporal community importance score was computed by summing their aggregated individual score in each year $t$.
Thus, the first edge-level feature at each timestamp $t$ in the training period, known as temporal community importance score for a keyword pair, can be computed as:

\begin{align}
\begin{split}
&score^H_t(a,b) = g^{a}_t\cdot v^{a}_t + g^{b}_t\cdot v^{b}_t
\end{split}
\label{eq:typology}
\end{align}
where $g^{a}_t$ and $v^{a}_t$ denote the genealogical community score and the corresponding centrality score of keyword $a$.
By considering the temporal sequences of individual community score and centrality values of both keywords,  we built a time series of this aggregated feature for each instance of non-connected keyword pair.
Since we considered three centrality values (i.e., $v^{au}_t$,  $v^{at}_t$ and $v^{d}_t$) for each keyword $v$ to determine its community score, the following notations will be used to denote three variants of this feature value:
$score^H_{au}(a,b)$ to denote the temporal community importance score for an edge $(a,b)$ where recursive centrality measures of keywords $a$ and $b$ were extracted from the keyword-author relations. 
Similarly, $score^H_{at}(a,b)$ will denote the same feature value where recursive centrality measures were extracted from the keyword-article relations.
Finally, $score^H_{d}(a,b)$ will be used to denote the same feature value where degree centrality values of keywords $a$ and $b$ were considered.

\subsubsection {Citation-weighted Recency}
\setlabel{Citation-weighted Recency}{subsec:cite_recency}	

An individual keyword's citation count represent its influence factor over other keywords.
However, domain-specific and representative keywords can achieve higher citation counts then the others.
Relative influence factor should also be brought into consideration.
For example, old keywords accumulate more citation over time then new keywords.
In contrast, some new keywords can gain significance quickly and acquire relatively high citation then other insignificant but old keywords.
Therefore, we considered the temporal factors of keyword's appearances such as current or distant appearances over time. 
Considering our training period (\eg, 2008-2014), an article published in 2008 gets more time to be cited than a article published in 2014. 
Despite, the later one carries more impact, the former one is  weighted more based on the numbers. 
Further, training year closer to the test period (e.g., 2014) is more significant than earlier years since recently co-appeared keywords are more likely to come together in near future, a fact known as `recency' in link prediction task ~\cite{yang2012predicting}.

Considering the aforementioned facts on temporal significance and citation count, we constructed our second edge-level feature for keyword pairs by accommodating both temporal recency and citation counts over time.
Since citation count and temporal recency are applicable to individual keywords, it requires to aggregate these counts for keyword-pairs.
For each keyword pair in year $t$, we aggregated their total citation counts. 
Further, we assigned temporal recency score depending on the appearances of both keywords in a particular year.
The recent their appearances are, or alternatively, the closer their co-appearances towards the test year, the higher the recency score.
Considering the keyword-pair $(a,b)$, let  $t$ denotes the training year $t \in [1, 2, 3, ... T]$, $h^u_t$ and $h^v_t$ denote the total number of citation counts of the articles in year $t$ where keywords $a$ and $b$ appears respectively, then the weighted-citation recency is calculated as follows:

\begin{align}
\begin{split}
&score^W_t(a,b) =  (h^a_t + h^b_t) \cdot \gamma t 
\end{split}
\label{eq:recency}
\end{align}
Here, $\gamma$ amplifies the recency effect and its value is two if both keywords in a keyword-pair appear in a year, one if either of the keyword appear, or zero (0) otherwise. 
$h^a_t$  will be assigned zero value if keyword $a$ does not appear in an article that received any citation.
However, the feature value will be augmented depending on the current appearances of the keywords (i.e., multiplied by value of $t$). 
The assumption here is that, keywords having high citation in the recent years will have high probability to appear together since they represent the recent trends.

\subsection {Feature Forecasting and Classification}
\setlabel{Feature forecasting and link classification}{sec:feature_forecast}

Considering temporal KCN $G_t(V_t,E_t)$, constructed for each year $t$ during the training period, we constructed time series of features for non-connected keywords pairs.  
Like other supervised dynamic link prediction study ~\cite{da2012time}, we employed deep-learning framework to forecast the future values of constructed features during the test phase (i.e., 2015).  
We use a Long Short Term Memory network ~\cite{hochreiter1997lstm} which is a special kind of Recurrent Neural Network (RNN) for both forecasting and classification tasks.
An LSTM takes sequential data as input and is considered well-suited in classifying temporal sequence. 
The LSTM used in this study consisted of two blocks of memory-cells with two different layers of hidden units. 
A simple LSTM cell unit takes three inputs ($X_t, h_{t-1}, C_{t-1}$). 
$X_t$ is the input of the current time step, $h_{t-1}$ is the output from the previous LSTM unit and $C_{t-1}$ is the “memory” of the previous unit. 
As for outputs, $h_t$ is the output of the current unit and $C_t$ is the memory of the current unit. 

In our experiment, time series of feature values was input into a 2-layer LSTM network. 
The number of timesteps for each training sample is seven since the training period is seven years (2008-2014) long. 
To forecast numerical features values, \textit{linear} activation function and MSE (Mean Squared Error) loss function were used.
Categorical features were encoded as one-hot vectors.
In this case, \textit{softmax} activation function and categorical cross-entropy (loss function) were used. 
In all cases, adam optimization technique was used. 
Further, these forecasted feature values were then fed as the training samples for our classification task. 
This classification task was implemented by adding another LSTM netwrok in the pipeline which includes a dense layer with a single neuron as output layer. 
A dense layer is a fully connected layer which means each neuron here receives  input from all the neurons in the previous layer, thus densely connected. 
In this layer, we used a logistic activation function named \textit{sigmoid} which is ideal for assisting in binary mutual exclusive classification problem. 
This output layer takes the forecasted feature values and predict the class (positive/negative) of the keyword pairs.   

\section {Results}
In this section we present our feature forecasting and link prediction results. 
We first present the feature forecasting performance followed by the classification performance demonstrated by the LSTM.
\subsection {Feature Forecasting Performance}
\setlabel{Feature Forecasting Performance}{subsec:forecast_performanse}
\begin{table}[!h]
\centering
\caption{Normalized RMSE values (0-1) calculated on  forecasted feature values against true feature values in the test year. Features include both node-level (keyword features) and edge-level (aggregated features for keyword-pairs). $score_{PA}(a,b)$ represents the preferential attachment score. $v^{at}_t$, and $v^{au}_t$ denote two recursive centrality values of keywords and $v^{d}_t$ denotes degree centrality values of keywords.}
\label{table:rmse_val}
\begin{tabular*}{\textwidth}{c@{\extracolsep{\fill}}cccccc}
\hline
Datasets & \multicolumn{3}{c}{$G_s$} & \multicolumn{3}{c}{$G_o$} \\ \hline
Number of Iteration & 100 & 500 & 1000 & 100 & 500 & 1000 \\ \hline
\multicolumn{7}{c}{Aggregated Features (Keyword pairs)} \\ \hline
$score^H_{at}(a,b)$  & 0.027 & 0.024 & 0.023 & 0.002 & 0.003 & 0.003 \\
$score^H_{au}(a,b)$  & 0.025 & 0.022 & 0.020 & 0.002 & 0.002 & 0.002 \\
$score^H_d(a,b)$  & 0.017 & 0.010 & 0.010 & 0.002 & 0.001 & 0.001 \\
$score^W_t(a,b)$ & 0.006 & 0.005 & 0.004 & 0.001 & 0.001 & 0.001 \\
$score_{PA}(a,b)$ & 0.016 & 0.015 & 0.015 & 0.005 & 0.004 & 0.004 \\ \hline
\multicolumn{7}{c}{Individual Features (Keyword)} \\ \hline
$v^{at}_t$ & 0.136 & 0.135 & 0.135 & 0.077 & 0.071 & 0.060 \\
$v^{au}_t$ & 0.119 & 0.112 & 0.110 & 0.070 & 0.063 & 0.057 \\
$v^{d}_t$ & 0.027 & 0.021 & 0.020 & 0.003 & 0.002 & 0.002 \\
Citation & 0.003 & 0.002 & 0.002 & 0.0004 & 0.0004 & 0.0004 \\ \hline
\end{tabular*}
\end{table}

We used LSTM to forecast the constructed feature values (\ref{subsec:feature_eng}) for the test period. 
For each keyword pair, time series of both keyword features (\ref{subsec:key_cen})  and aggregated features (\ref{subsec:heredity} and \ref{subsec:cite_recency}), computed during the training period, were input in LSTM.
To evaluate the performance of this forecasting model, we computed the RMSE (root mean squared error) values by considering the actual feature values in the test year against the forecasted values.
Table  \ref{table:rmse_val} present the RMSE values to measure the performance of LSTM in forecasting.
We trained the LSTM model for different number of iterations (i.e. epochs). 
The effect of iteration number is visible in the rmse values.
In most cases, higher number of iteration results in more robust model training. 
The range of actual values for different features can be varied significantly. 
So we presented the normalized RMSE values in the range of 0-1. 
RMSE values in $G_o$ denote that in this dataset, the forecasting errors were smaller than those in $G_s$ which can be attributed to the size of the dataset. 
In this table, forecasting performance for both the keyword feature and aggregated features are presented. 
For comparison's sake, we also calculated RMSE in forecasting the  \textit{preferential attachment} score which is a widely used metric in link prediction task.
It is noteworthy that preferential attachment is an aggregated network feature which is computed by multiplying the number of neighbours of each keyword in a keyword pair.
\begin{figure}[!h]
    \centering
    \includegraphics[width=.9\textwidth]{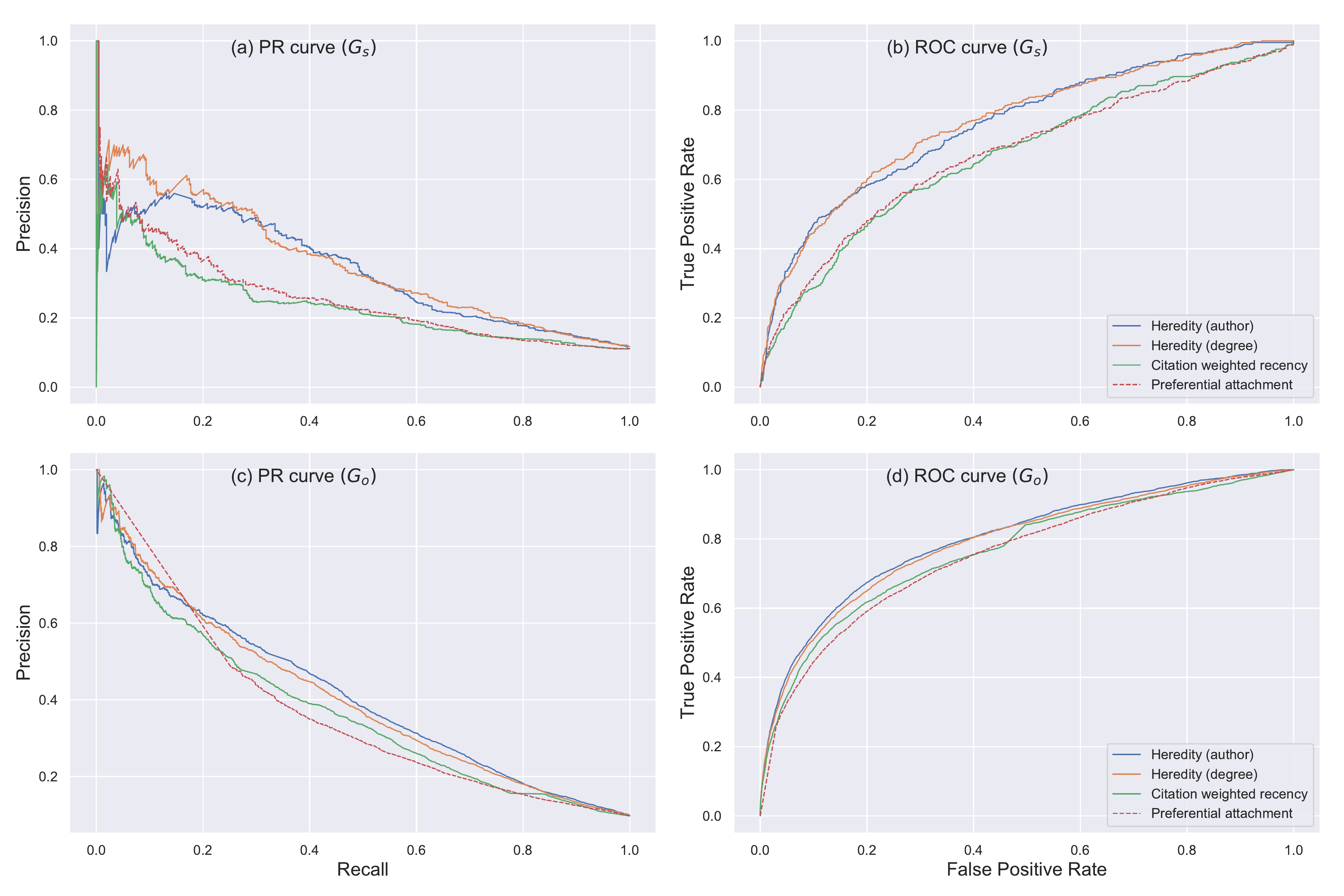}
    \caption{P-R and ROC curves in both datasets ($G_s$ and $G_o$) to demonstrate the classification performance of the LSTM classifier using the feature values constructed in this study. Traditional topological similarity metric `Preferential Attachment' was also used to compare the performance with the constructed features. `Heredity (author)' denotes the aggregated feature temporal community importance $score^H_{au}(a,b)$. Similarly,`Heredity (article)' denotes the temporal community importance $score^H_{at}(a,b)$. }
    \label{fig:pr_curve}
\end{figure}
\subsection{Link Prediction Performance}
\setlabel{Link Prediction Performance}{subsec:link_performance}

The supervised link prediction framework is subject to highly imbalanced data with a very large number of instances with negative label. 
In practice, only a few pairs of keywords participate in true emerging edges out of every possible pairs.
Therefore, following other studies like in ~\cite{choudhury2016time}, the ratio of positive and negative class instances was set to 1:10.
We also used 30\% of the instances for validation and evaluation purposes.  
Similar to forecasting performance, $G_{o}$ dataset was found to demonstrate high performance in comparison to $G_{s}$.
The AUCROC (i.e., Area Under ROC Curve) is commonly used for evaluating such imbalanced classification problems.
It introduces a probability value to quantify the uncertainty associated with the classifiers. In case of binary classification, AUCROC enforces larger weight on smaller class by using this threshold value. 
In Table ~\ref{table:auc_val}, we present both the accuracy and AUCROC scores computed in both $G_{s}$ and $G_{o}$ datasets. 
For comparison's sake, we also present the performance of \textit{\textbf{Preferential Attachment}} metric which is a well-known topological similarity metric widely used in supervised link prediction task.
In comparison to this topological similarity metric, it is evident from Table \ref{table:auc_val} that the features constructed in this study outperformed the traditional and widely prevalent metric for link prediction in KCNs, the preferential attachment.
Better performance was observed in $G_o$ rather than $G_s$ which can be attributed to the greater number of instances.
It is also evident that the first aggregated feature (temporal community importance) performed better than the  citation weighted recency scores. 

\begin{table}[H]
\centering
\begin{threeparttable}
\caption{Dynamic supervised link prediction performance using AUCROC (AUC) and Accuracy (Acc\%) values.}
\label{table:auc_val}
\begin{tabular*}{\textwidth}{l@{\extracolsep{\fill}}rrrr}
\toprule
{} &  \multicolumn{2}{c} {$G_{s}$} & \multicolumn{2}{c} {$G_{o}$}\\
\midrule
{} &  AUC &  Acc(\%) & AUC & Acc(\%)\\
\midrule

$score^H_{at}(a,b)$    					&   0.749 & 89.3  	&    0.783 	& 90.9\\
$score^H_{au}(a,b)$   					&   0.759 & 88.7 	&    0.800 	& 91.0\\
$score^H_{d}(a,b)$    					&   0.765 & 89.5  	&    0.794 	& 91.0\\
$score^W_t(a,b)$  				&   0.680 & 88.9  	&    0.776 	& 90.7\\
$score_{PA}(a,b)$      			&   0.698 & 88.8  	&  	 0.753 	& 90.2\\
\bottomrule
\end{tabular*}
\end{threeparttable}
\end{table}

Considering other evaluation metrics, the P-R (precision - recall) curve depicts the precision-recall trade-off for a classifier.
This measurement is widely used in information retrieval.
Reviewing both precision and recall is useful in cases where there is an imbalance in the instances between the two classes.
The reason for this is that typically the large number of negative class instances means we are less interested in the skill of the model at predicting negative class instances correctly, (i.e., high true negatives).
The most crucial thing of P-R curve is that this calculation does not make use of the true negatives. 
It is only concerned with the correct prediction of the minority class, the positive class instances.
A P-R curve is a plot of the precision in y-axis and the recall in x-axis.
In ROC curve, the goal is to have a model be at the upper left corner, which is basically getting no false positives.
Whereas, in P-R curve, the goal is to have a model be at the upper right corner, which is basically getting only the true positives with no false positives and no false negatives.
In Figure \ref{fig:pr_curve}, we present both the ROC and P-R curves of our features (solid lines) including the Preferential Attachment (dashed lines) metrics.
The top row represents P-R and ROC curves in $G_s$ and the bottom row represents the same in $G_o$.
Like the aforementioned performance measurements, in most cases of this figure, the features constructed in this study outperformed the preferential attachment metric in both datasets.

\subsection {Distribution of Aggregated Feature Values}
\begin{figure}[!h]
    \centering
    \includegraphics[width=1\textwidth]{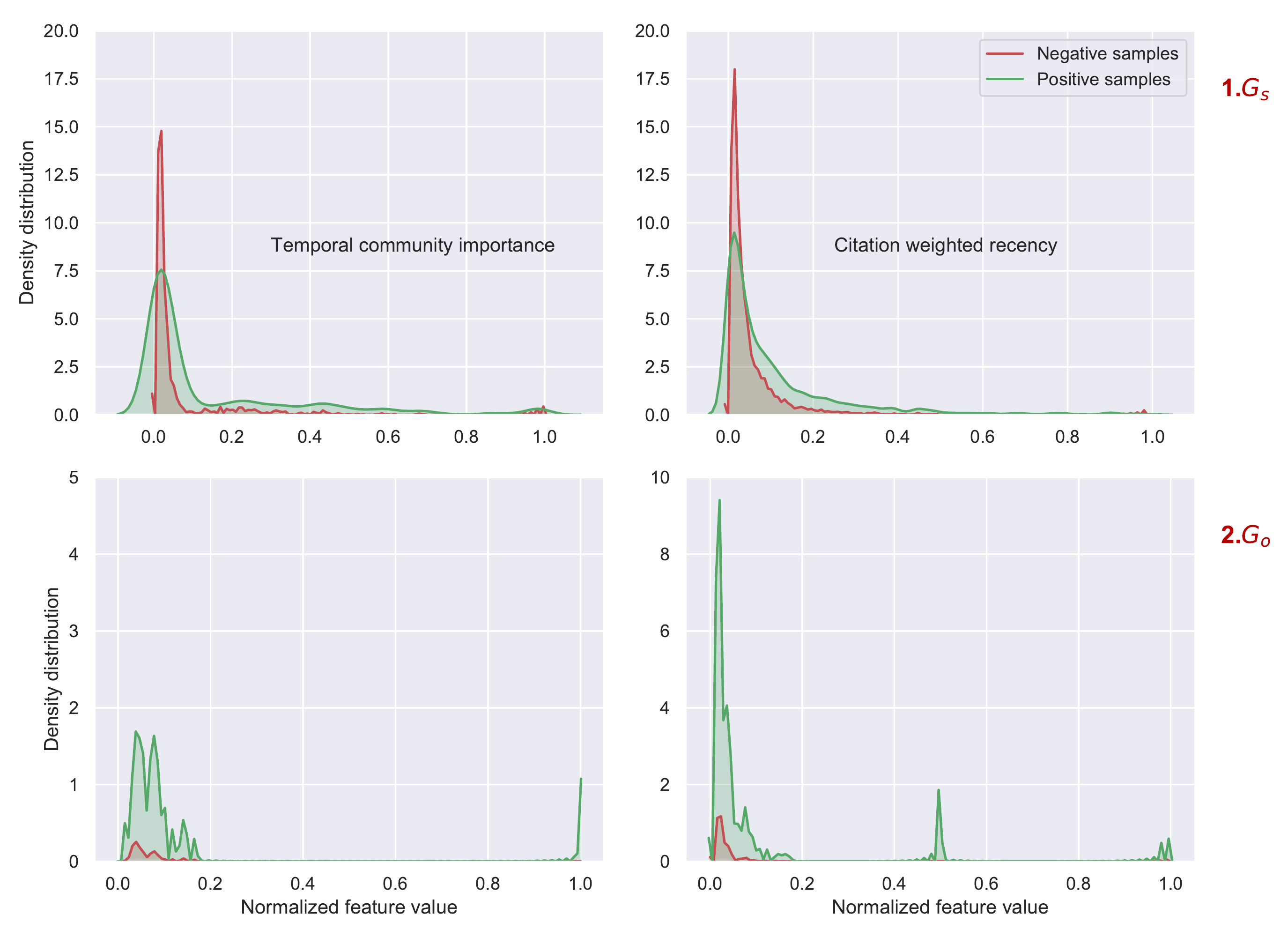}
    \caption{ Positive and negative class density of two aggregated features: temporal community importance score  $score^H_{au}(a,b)$ where the recursive centrality measures were extracted from the keyword-author relations and Citation weighted recency $score^W_t(a,b)$ in both datasets }
    \label{fig:distribution}
\end{figure}

\setlabel{Distribution of feature values}{subsec:dist_feature}
In Figure ~\ref{fig:distribution}, we present the distributions of feature values for the constructed aggregated features (i.e., temporal community importance $score^H_{au}(a,b)$ and citation weighted recency).
For all positive and negatively labelled keyword pairs (samples), the normalized feature values for both aggregated features are presented as kernel density plots.
Distribution of the feature values for the positive samples (true keyword pairs in the test year) are presented in green color and the negative samples are presented in red color.
It is noteworthy that, for temporal community importance score, we only consider the recursive centrality values extracted from the keyword-author relations.

\begin{figure}[!h]
    \centering
    \includegraphics[width=1\textwidth]{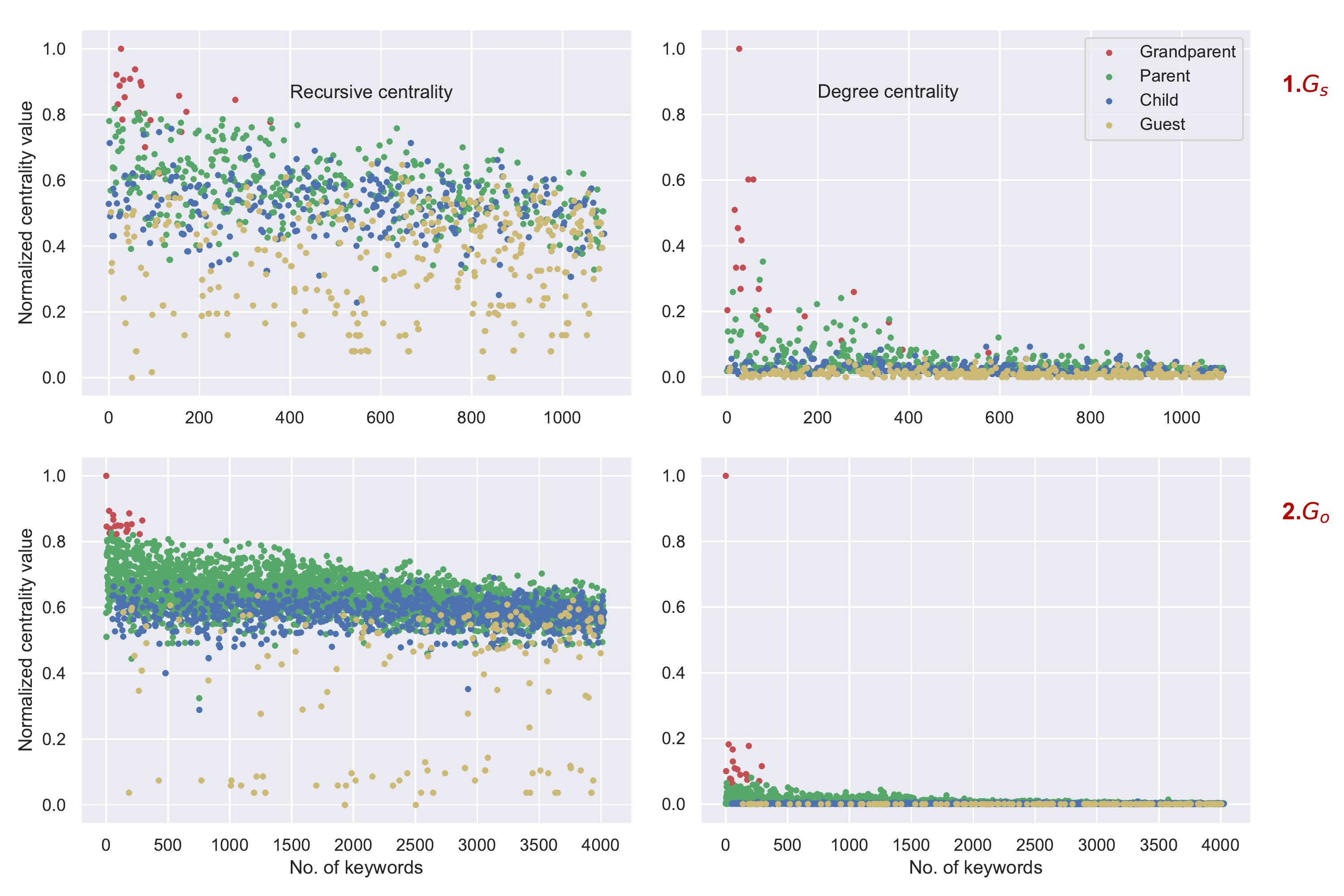}
    \caption{ Two different centrality values (i) recursive centrality values extracted from the keyword-author relations (left) and (ii) degree centrality  of the keywords participating in the positively labeled edges during the test year in both datasets.  }
    \label{fig:variance}
\end{figure}
Two observations are evident from these figures. 
Firstly, since the overlap between the red and green colored regions may trigger classification errors, the lower the overlap, the higher the classification performance. 
The amount of overlap signifies that the aggregated features constructed in this study have non-trivial discriminatory characteristics.
Secondly, in contrast to topological similarity metrics where high value of the metrics corresponds to higher similarity between a pair of nodes (\eg, having more common neighbours between two nodes in a network), we found that positively labelled keyword pairs have high density in lower feature values.
This denotes that instead of higher values of the features, comparatively lower features values have high probability in forming emerging relationships.
Although, in $G_o$, we observe high feature values for positively labeled edge instance in the test period.
This fact is contrary to our initial assumption, mentioned in \ref{subsec:feature_eng} that higher values denote higher probability of keyword co-occurrences.
Although, exploring the reason behind this is out of the scope of this study, however, in Figure \ref{fig:variance}, we present the normalized keyword feature values (recursive centralities and degree centrality) computed for the test year (i.e., 2015) in both datasets.
The lower feature values can be attributed to the number of children and guest keywords including the variances in normalized centrality values of the parent keywords.
It is noteworthy that from the aforementioned figures and tables, we have observed that the parent community dominates the merging link formations.

\section {Discussion and Conclusion}
Scientific progress depends on formulating verifiable and deductible hypotheses generation. 
This requires both understanding and informed inferences from existing knowledge and information ~\cite{choi2018literature}.
Rapid growth of scientific knowledge and over specializations (domain-specific fragmentation) may engender opportunities to derive solutions from one domain to address problems in another, although the underlying relationship may remain implicit or the concerned groups from both domains are unaware of the work of each other ~\cite{hristovski2001literature}. 
However, the continuous surge in published scientific literature limits the scope of analyses an individual can accomplish to extract these novel and implicit relationships between disjoint concepts, topics and domains ~\cite{wren2004knowledge}.
Proliferation of scientific production inhibits scientists and policy makers to detect trending subject areas and the linkages among these areas in their research fields, and mapping the dynamics of science to plan for research progress ~\cite{he1999knowledge}.
human comprehension of such massive information and knowledge is challenging when it exceeds the scale of human analytical skills.
For example, as mentioned by Spangler et al. ~\cite{spangler2014automated}, it is inconceivable for a scholar to possibly assimilate, recall and accurately process all the known facts relevant to protein functions, relationships between proteins and identifications of roles of a particular protein related to a disease while there are over 70000 articles on a single protein - the tumour suppressor 'p53`.
Thus, there is a great difference between what is known, and what we know as individuals from the collective and multidisciplinary knowledge within a given domain. 
Further, the specialization  or fragmentation of literature may promote poor communication between specialties since scientists tend to communicate more within their fragments than the broader community engaged within the domain ~\cite{swanson2001asist}. 
According to Ganiz et al.  ~\cite{ganiz2005recent} Literature Based Discovery (LBD) addresses the challenge of  seemingly boundless increases in scientific knowledge including knowledge overspecialization faced by the scientific communities.
However, existing LBD models suffers from the lack of generalized predictive model to successfully predict the emerging trends in such discoveries.
Despite their success, LBD techniques including text analysis, information retrieval and natural language processing are deprived of the benefits of bibliometrics, specially the analytical advantages of temporal keyword co-occurrence networks (KCNs).
These KCNs and network analysis methods are found to be supportive in identifying technological trends, analyse research topics and follow their evolution and track the development of innovation system research ~\cite{dotsika2017identifying}.
Further, temporal dynamics of KCN and community-aware features are underutilized in the process of literature-based hypotheses generation.
To this end, this study proposed a predictive framework that integrates temporal evolution of KCN, genealogical communities, citation counts of keywords, and an LSTM based forecasting and prediction model.
The KCNs in this study are comprised on author selected keywords which best describe the research themes referred by the corresponding authors and are also considered as knowledge entities.
This framework includes feature engineering process that builds novel features for both keywords (node) and keyword pairs (edge defined by co-occurrence).
In this study, we developed two recursive centrality measures by considering two types of bipartite relations: keyword-author and keyword-article.
We also considered the degree centrality of keywords extracted from temporal unipartite KCNs.
These centrality measures were used to define temporal communities based genealogical relationships among keywords. 
Temporal citation counts of keywords were also used as keyword features.
Bipartite recursive centrality values including the degree centrality, genealogical community information and relative importance of temporal citation counts were used to construct edge-level (keyword pairs) features.
Seven years of training period was used to built time series of both node and edge-level features those were input into an LSTM network to forecast the feature values in the test year.
If we forecasted only the edge-level features, then the individual node-level significance of keywords would have lost due to the aggregation of mixed data types (eg. numerical, categorical) in edge-level.
For example, in case of the temporal community importance score, originating from the recursive centrality values, extracted from keyword-author bipartite relations, the feature vector includes both centrality values of two keywords as well as their corresponding community and the aggregated feature value.
The centrality values are numerical, however, the community is categorical which was represented as one-hot vector.
This can easily be handled in LSTM instead of other forecasting technique.
With relatively trivial forecasting error, the forecasted feature values were used in supervised link prediction to classify both positive and negatively labeled non-connected keyword pairs.
The performance of the LSTM classifier was measure using well-known performance metrics and also compared against well-known network topological similarity metric used in link prediction.
High performance of the constructed feature indicates that these features are not only supportive in dynamic supervised link prediction but also can be beneficial in predicting literature based knowledge discovery or emerging trend detection.

Despite the better performance measurements, this study is not free from its limitation.
The first limitation comes from the lack of accommodating domain experts in identifying semantic similarity of keywords and various forms of abbreviations.
Secondly, in this study we used manual name disambiguation instead of any standard methods to clean the author names including which can further be explored in future studies.
The semantic relationships between textual contents were ignored in this study.
The reason behind is that we only considered the author selected keywords those are not semantically presented in the scientific literature but only co-appears if the corresponding authors recognize them as relevant. 
Finally,lack of any such deep learning-based predictive framework in literature based discovery to compare the performance of the proposed framework instead of relying on tradition link prediction metric.
This study can further be extended in various ways.
Firstly, this study considered previous year's information to identify communities of keywords in the current year.
However, future studies can consider more historical information can be used (more than one previous year information) to determine such communities.
Other centrality measures and network community detection algorithm can used to compare the performance of link prediction using community-aware features.
Further, some weighting strategies can be followed to assign weights on edges between keywords belonging to either same community or different communities. 



\section{References}
\bibliographystyle{elsarticle-num} 
\bibliography{main.bib}      



\section*{Author contributions}

N.C. conceived the manuscript idea, collected data, design experiments and contributed in writing, and F.F cleaned the datasets, conducted the experiment(s), developed features, analysed the results, and wrote the research methodology section. M.K. contributed in writing and reviewed the manuscript. 

\section{Dataset and Code}
The dataset and programming codes used in this study is available a https://github.com/faisal-iut/linkPrediction

\end{document}